\documentclass[twocolumn,prb,10pt,aps,floatfix]{revtex4-1}
\usepackage[english]{babel}
\usepackage{graphicx}
\usepackage[pdftex]{epsfig}
\usepackage{ragged2e}
\usepackage{amsmath,amssymb}
\usepackage{times}
\usepackage[colorlinks,linkcolor=blue,citecolor=blue,urlcolor=blue]{hyperref}
\usepackage{color}
\usepackage[table,xcdraw]{xcolor}
\usepackage{epstopdf}
\usepackage{physics}
\usepackage[export]{adjustbox}
\usepackage{dsfont}

\newcommand{\be}{\begin{equation}}
\newcommand{\ee}{\end{equation}}

\newcommand{\bee}{\begin{equation*}}
\newcommand{\eee}{\end{equation*}}

\newcommand{\rr}{{\boldsymbol r}}
\newcommand{\rs}{{{\boldsymbol r}'}}
\newcommand{\rrs}{{{\boldsymbol r}{\boldsymbol r}'}}

\newcommand{\dr}{{\delta{\boldsymbol r}}}

\newcommand{\pair}{{({\boldsymbol r},{\boldsymbol r}')}}
\newcommand{\T}{\mathcal{T}}
\newcommand{\drr}{{\delta{\boldsymbol r}}}
\newcommand{\dx}{{\delta x}}
\newcommand{\dy}{{\delta y}}
\newcommand{\kk}{{\boldsymbol k}}

\newcommand{\drs}{\includegraphics[scale=1]{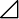}}
\newcommand{\dls}{\includegraphics[scale=1]{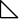}}
\newcommand{\uls}{\includegraphics[scale=1]{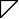}}
\newcommand{\urs}{\includegraphics[scale=1]{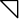}}

\newcommand{\drb}{\includegraphics[scale=1.3]{dr.pdf}}
\newcommand{\dlb}{\includegraphics[scale=1.3]{dl.pdf}}
\newcommand{\ulb}{\includegraphics[scale=1.3]{ul.pdf}}
\newcommand{\urb}{\includegraphics[scale=1.3]{ur.pdf}}

\begin{document}
\title{Topological spinon bands and vison excitations in spin-orbit coupled quantum spin liquids}
\author{Jonas Sonnenschein$^{1}$}
\author{Johannes Reuther$^{1, 2}$}
\affiliation{$^1$Dahlem Center for Complex Quantum Systems and Institut f\"ur Theoretische Physik, Freie Universit\"{a}t Berlin, Arnimallee 14, 14195 Berlin, Germany}
\affiliation{$^2$Helmholtz-Zentrum f\"{u}r Materialien und Energie, Hahn-Meitner-Platz 1, 14019 Berlin, Germany} 

\date{\today}

\begin{abstract}
Spin liquids are exotic quantum states characterized by the existence of fractional and deconfined quasiparticle excitations, referred to as spinons and visons. Their fractional nature establishes topological properties such as a protected ground-state degeneracy. This work investigates spin-orbit coupled spin liquids where, additionally, topology enters via non-trivial band structures of the spinons. We revisit the $\mathds{Z}_2$ spin-liquid phases that have recently been identified in a projective symmetry-group analysis on the square lattice when spin-rotation symmetry is maximally lifted [Phys. Rev. B 90, 174417 (2014)]. We find that in the case of nearest neighbor couplings only, $\mathds{Z}_2$ spin liquids on the square lattice always exhibit trivial spinon bands. Adding second neighbor terms, the simplest projective symmetry-group solution closely resembles the Bernevig-Hughes-Zhang model for topological insulators. Assuming that the emergent gauge fields are static we investigate vison excitations, which we confirm to be deconfined in all investigated spin phases. Particularly, if the spinon bands are topological, the spinons and visons form bound states consisting of several spinon-Majorana zero modes coupling to one vison. The existence of such zero modes follows from an exact mapping between these spin phases and topological $p+ip$ superconductors with vortices. We propose experimental probes to detect such states in real materials.
\end{abstract}
\maketitle

%====================================================================
\section{Introduction}\label{introduction}
Quantum spin liquids are fascinating spin phases that evade classical magnetic ordering in the ground state due to magnetic frustration effects.\cite{balents10, savary16,lee08} While originally, these states have been described within the resonating-valence bond (RVB) paradigm by P. Anderson,\cite{anderson73} the past decades have unveiled that their internal structure is much more complex than the picture of fluctuating singlet dimers might suggest. In particular, spin liquids exhibit various characteristic topological properties which manifest themselves in fractional spin excitations and a topologically protected ground-state degeneracy.\cite{wen_book,frustration_book} In the simplest case, these properties follow from an underlying $\mathds{Z}_2$ gauge theory\cite{wegner71,kogut79,senthil00} where fractional spinon excitations\cite{read91,wen91} (that effectively behave as half of a conventional spin-1 spin-flip operation) couple to fluctuating $\mathds{Z}_2$ gauge fields. Together with flux excitations of the gauge field (also referred to as visons)\cite{read89, kivelson89,punk14} the spinons represent the fundamental quasiparticles of a spin liquid. While the visons are spinless bosonic particles, a $\mathds{Z}_2$ gauge theory can be formulated for both, fermionic and bosonic spinons.

An appealing feature of a $\mathds{Z}_2$ gauge-theory description for quantum spin liquids is that spinons (as well as visons) naturally appear as deconfined particles that do not experience any long-range confining forces. This is in contrast to U(1) gauge theories where the effects of gauge fluctuations may destabilize spin-liquid phases, possibly driving the systems into conventional long-range magnetically ordered states.\cite{polyakov77,hermele04,ran09} The deconfined property of spinon excitations opens up the striking possibility of effectively realizing the physical phenomena of free-fermion systems -- including different types of band structures or superconductivity -- in the extreme opposite limit of strongly correlated Mott insulators. (We note that spinon superconducting pairing terms are a necessary condition to achieve a $\mathds{Z}_2$ gauge structure.\cite{wen91}) This idea becomes particularly interesting when effects of spin-orbit coupling are included as this allows for the formation of non-trivial band topologies in the spinons, effectively leading to ``topological spinon insulators''.\cite{pesin10,ruegg12,rachel10,cho12} In a way, such states may be considered as ``double topological'' in the sense that quasiparticles which as are already topological (fractional) in the first place, additionally exhibit a topological band structure.

While these phases nicely combine physical concepts of two extreme limits -- strong coupling versus weak coupling -- they are, unfortunately, extremely hard to investigate on the level of actual model systems (and it is probably even harder to find material realizations). This is mainly because for a generic spin Hamiltonian there is presently no numerical approach available which directly probes fractional spin-liquid excitations and their dynamical properties. To still gain insight into the properties of the aforementioned states, mainly two different strategies are currently pursued. First, Kitaev spin models on tri-coordinated lattices\cite{kitaev06} allow for an exact analytical solution of the spectrum of fermionic ``matter fields'' (which effectively take over the role of spinons but are typically described in terms of Majorana fermions) and flux excitations.\cite{baskaran07,knolle14,brien16} In these systems, various interesting phenomena such as topological Majorana band structures,\cite{kells10,thakurathi14,hermanns15,brien16} bound states between Majorana fermions and fluxes,\cite{knolle15,rachel16,theveniaut17} or the nucleation of Majorana bands in flux-superlattices\cite{lahtinen12,lahtinen14} have been investigated. However, to guarantee exact solvability one is restricted to models with specially designed Ising interactions and perturbations beyond these couplings typically complicate their analysis enormously. 

A second possibility for investigating fractional excitations in quantum spin liquids relies on the more general concept of the ``projective symmetry group'' (PSG) method\cite{wen02} that is applied in this work. This approach does not start with a specific spin Hamiltonian but rather assumes a certain set of symmetries of the system (e.g. lattice symmetries and time reversal invariance). Using a fermionic parton representation for spin operators,\cite{abrikosov65} the interaction terms are mean-field decoupled in all possible symmetry-allowed ways, leading to a systematic classification of spin liquid phases characterized by different types of free spinon band structures (note that the PSG approach may also be formulated in terms of Schwinger bosons instead of fermions\cite{wang06,wang10,messio13}). So far the PSG method has mostly been applied to Heisenberg models on different two dimensional lattices,\cite{wang06,chen12,messio13,lu11,wang10,lu11_2,messio12,bieri16,schaffer17} where -- depending on the precise set of symmetries -- hundreds of different spin-liquid phases are identified. In contrast, the PSG approach is rarely applied to spin-orbit coupled systems with anisotropic spin interactions.\cite{schaffer13,dodds13,huang17,reuther14} One such analysis recently classified all possible $\mathds{Z}_2$ spin liquids on the square lattice when spin rotation symmetry is maximally lifted and identified as many as 1760 solutions.\cite{reuther14} Remarkably, it has been found that topological $p+ip$ superconducting pairing represents a generic property of the spinons which, in many cases, leads to topologically non-trivial band structures.

In this paper, we revisit the PSG classification of spin-orbit coupled spin liquids on the square lattice and investigate their properties in a more realistic and simplified setting. Indeed, the large number of states quoted above mainly stems from the rather unrealistic assumption that interactions occur for all possible bond distances. By restricting the couplings to first or second neighbor interactions on the square lattice, the number of spin-liquid phases reduces drastically such that the most generic states may be identified and studied. We find, however, that with nearest neighbor interactions only, spinon band structures of $\mathds{Z}_2$ spin liquids are always topologically trivial, which can be traced back to the projective implementation of time-reversal symmetry. Once second neighbor couplings are added, topological spinon bands appear even under the most simplifying assumptions and we find that the generic spinon Hamiltonian resembles the Bernevig-Hughes-Zhang (BHZ) model\cite{bernevig06,konig08} for the quantum spin Hall material HgTe.\cite{konig07}

A particular focus of this work is on vison excitations. Using an approximation that treats visons as static quasiparticles, their interaction potential turns out to be of extreme short-range nature, confirming that they are effectively deconfined. Remarkably, depending on the Chern numbers of the topological spinon bands we observe that visons can bind multiple zero modes from the spinon sector. For topological spinon bands and in the limit of large vison distances, the excitation energies of such states decay exponentially, indicating that the bound spinons are effectively described by Majorana zero modes. The connection between topological spinon bands and Majorana bound states\cite{slager15} is made explicit by establishing an exact mapping of our spin-liquid quasiparticles onto a topological $p+ip$ superconductor with quantum vortices. We finally consider the more realistic situation where visons form a dilute gas on randomly arranged local defects, e.g., mimicking the effects of finite temperatures. Due to the tight coupling between Majorana modes and visons, the bound states form a narrow band around zero energy if the system is in a topologically non-trivial phase. With increasing vison density this band is populated by an increasing number of states. We conclude that such effects might allow one to experimentally identify topological spinon band structures.

The paper is organized as follows: In Section~\ref{psg} we review the PSG approach for spin systems with maximally lifted spin-rotation symmetry as has been applied in Ref.~\onlinecite{reuther14}. We identify the simplest of such PSG solutions in Section~\ref{short-range} by first restricting to nearest neighbor couplings (Section~\ref{nearest}) and then adding second neighbor terms (Section~\ref{sec_neighbors}). Particularly, we discuss the spinon band structures and topological phase diagrams of three selected spin-liquid phases. In the following Section~\ref{visons} we study the properties of vison excitations in these states. After reviewing some basic concepts of $\mathds{Z}_2$ lattice gauge theories in Section~\ref{review_gauge_theory}, we determine the effective vison pair potentials in Secion~\ref{deconfinement}, confirming that visons are deconfined. Section~\ref{bound_states} investigates spinon-vison bound states and formulates a mapping of our BHZ-like spin liquid to a topological $p+ip$ superconductor with vortices. Finally, Section~\ref{vison_gas} addresses the properties of a gas of randomly arranged visons. The paper ends with a conclusion and outlook in Section~\ref{conclusions}.

%====================================================================
\section{PSG classification of spin liquids with spin-orbit coupling}
\label{psg}
\subsection{Mean-field decoupling}
Before we investigate specific spin liquid states in the next section, we first briefly review the PSG classification procedure in the case of spin-anisotropic systems. For more in-depth discussion we refer the interested reader to Refs.~\onlinecite{wen02,reuther14}. The starting point of the PSG analysis is an anisotropic two-body spin Hamiltonian with the general form
\begin{equation}
H=\sum_{(\rrs)} J_{\rrs}^{ij}S_\rr^{i} S_\rs^j\;,\label{ham}
\end{equation}
where $S^{i}_{\rr}$ denotes the $i$th component ($i=1,2,3$) of a spin-1/2 operator at lattice position $\rr$ and $J_{\rrs}^{ij}$ are the exchange couplings. The sum runs over pairs of sites as indicated by the symbol $(\rrs)$. Note that repeated indices $i$, $j$ are implicitly summed over. We will not further specify the couplings $J_{\rrs}^{ij}$ but assume that the Hamiltonian respects all lattice symmetries of the square lattice. Through their dependence on the components $i$ and $j$, the interactions may, however, break continuous spin-rotation symmetries.

We apply a fermionic version of the PSG approach where the spin operators are written in terms of parton operators,\cite{abrikosov65}
\be
S_\rr^j=\frac{1}{2}f_\rr^\dagger \sigma^j f_\rr\;.
\ee
Here, $\sigma^{j}$ are the Pauli matrices and $f_\rr=(f_{\rr\uparrow},f_{\rr\downarrow})^\text{T}$ denotes a two-component spinor of fermionic annihilation operators $f_{\rr\alpha}$ with $\alpha=\uparrow,\downarrow$. The parton representation doubles the dimension of the local Hilbert space on each site where the physical spin-1/2 states are those that satisfy the single occupancy constraint $\sum_\alpha f_{\rr\alpha}^\dagger f_{\rr\alpha}=1$ or equivalently $f_{\rr\uparrow}f_{\rr\downarrow}=0$. These conditions may also be expressed as a gauge freedom, according to which the states in the physical sector of the Hilbert space are those that remain unaffected by the local gauge transformation
\be
\psi_\rr\rightarrow w_\rr \psi_\rr\;,\label{gauge_trafo1}
\ee 
where $\psi_\rr=(f_{\rr\uparrow},f_{\rr\downarrow}^\dagger)^\text{T}$ is a two-component spinor in Nambu space and $w_\rr$ is an arbitrary (site-dependent) $2\times2$ SU(2) matrix obeying $w_\rr^\dagger=w_\rr^{-1}$.

Next, the fermionic version of Eq.~(\ref{ham}) is mean-field decoupled in all hopping and pairing channels as described by the mean-field amplitudes $\langle f_{\rr\alpha}^\dagger f_{\rs\beta}\rangle$ and $\langle f_{\rr\alpha} f_{\rs\beta}\rangle$, respectively (note that no decoupling is performed in the local channel $~\langle f_{\rr\alpha}^\dagger \sigma^{j}_{\alpha\beta} f_{\rr\beta}\rangle$ as this would describe trivial, classical magnetically ordered states). In the general case considered here, the original spin Hamiltonian~(\ref{ham}) may break all continuous spin-rotation symmetries such that the mean-field procedure formally generates anisotropic amplitudes with all possible combinations of $\alpha$ and $\beta$ in spin space. In compact notation, the decoupled Hamiltonian $H_\text{mf}$ can be written in terms of the full four-component Nambu spinor $\Psi_\rr=(f_{\rr\uparrow},f_{\rr\downarrow}^\dagger,f_{\rr\downarrow},-f_{\rr\uparrow}^\dagger)^\text{T}$, yielding
\be
H_{\text{mf}}=\sum_\pair\left( \Psi_\rr^\dagger U_\rrs\Psi_\rs+\text{H.c.}\right)\;,\label{ham_mf}
\ee
with the $4\times4$ matrix $U_\rrs$ containing all mean-field amplitudes. For a PSG analysis it is convenient to write $U_\rrs$ in the form
\be
U_\rrs=
\left(\begin{array}{cc}
u_\rrs^s+u_\rrs^{t_1} & u_\rrs^{t_2}+u_\rrs^{t_3} \\
-u_\rrs^{t_2}+u_\rrs^{t_3} & u_\rrs^s-u_\rrs^{t_1} 
\end{array}\right)\label{block}
\ee
where each entry is a $2\times2$ matrix. These blocks can be expanded in terms of Pauli matrices and the identity matrix $\sigma^0$,
\begin{eqnarray}
u^s_\rrs=i s^0_\rrs\sigma^0+s^j_\rrs \sigma^j\;,
u^{t_1}_\rrs=t^0_{1,\rrs}\sigma^0+it^j_{1,\rrs}\sigma^j\;,\notag\\
u^{t_2}_\rrs=it^0_{2,\rrs}\sigma^0+t^j_{2,\rrs} \sigma^j\;,
u^{t_3}_\rrs=t^0_{3,\rrs}\sigma^0+i t^j_{3,\rrs} \sigma^j\;.\notag\\
\end{eqnarray}
Here, the coefficients $s^j_\rrs$ and $t_{1/2/3,\rrs}^j$ are real mean-field amplitudes. In Eq.~(\ref{block}) the different entries can be distinguished according to their behavior under spin rotations. The term $u_\rrs^s$ is spin-isotropic and describes (spin-independent) hopping $f^\dagger_{\rr\uparrow}f_{\rs\uparrow}+f^\dagger_{\rr\downarrow}f_{\rs\downarrow}$ and singlet pairing $f_{\rr\uparrow}f_{\rs\downarrow}-f_{\rr\downarrow}f_{\rs\uparrow}$. All other matrices $u_\rrs^{t_{1/2/3}}$ contain fermionic bilinears in the triplet channel, i.e., their action is associated with a spin flip along a certain direction. Particularly, the term $u_\rrs^{t_1}$ breaks SU(2) spin rotation symmetry down to U(1) symmetry for rotations around the $z$-axis. The corresponding mean-field amplitudes describe spin-dependent hopping $f^\dagger_{\rr\uparrow}f_{\rs\uparrow}-f^\dagger_{\rr\downarrow}f_{\rs\downarrow}$ and triplet pairing $f_{\rr\uparrow}f_{\rs\downarrow}+f_{\rr\downarrow}f_{\rs\uparrow}$. Finally, the matrices $u_\rrs^{t_2}$ and $u_\rrs^{t_3}$ represent spin-flip hopping $f^\dagger_{\rr\uparrow(\downarrow)}f_{\rs\downarrow(\uparrow)}$ and spin-polarized triplet pairing $f_{\rr\uparrow(\downarrow)}f_{\rs\uparrow(\downarrow)}$ which also break the remaining U(1) symmetry.

In analogy to Eq.~(\ref{gauge_trafo1}), the gauge transformation can be formulated for the four-component spinor $\Psi_\rr$, yielding
\be
\Psi_\rr\rightarrow W_\rr\Psi_\rr\quad\text{with}\quad W_\rr=
\left(\begin{array}{cc}
w_\rr & 0 \\
0 & w_\rr
\end{array}\right)\;,\label{gauge_trafo2}
\ee
where $w_\rr$ is a $2\times2$ SU(2) matrix.
 
The obvious benefit of a mean-field decoupling is that $H_\text{mf}$ describes free fermions and can be solved exactly. Furthermore, the fermions can be naturally associated with spinons which are deconfined and fractional quasiparticles in a spin liquid. Hence, a free fermionic model represents a good starting point for describing spin liquids and their emergent spinon excitations but still lacks the correct gauge structure. This can be seen by gauge-transforming the mean-field Hamiltonian $H_\text{mf}$ in Eq.~(\ref{ham_mf}) using site-dependent matrices $W_\rr$ which effectively changes the mean-field matrix $U_\rrs$ according to $U_\rrs\rightarrow W^\dagger_\rr U_\rrs W_\rs$. In the generic case, $U_\rrs\neq W^\dagger_\rr U_\rrs W_\rs$ showing that the mean-field Hamiltonian does not fulfill the local gauge invariance of the original Hamiltonian. This indicates that $H_\text{mf}$ also operates in the {\it unphysical} sector of the Hilbert space and that its eigenstates are not even proper spin states obeying the parton constraint. As explained in Section~\ref{visons} a gauge invariance can be restored by allowing for {\it fluctuating} amplitudes $U_\rrs$, resulting in an effective gauge-theory description with additional gauge-field degrees of freedom and vison quasiparticles. The structure of such fluctuations [$\mathds{Z}_2$, U(1), ...] can already be determined on the bare mean-field level and is connected to the concept of the so-called {\it invariant gauge group} (IGG).\cite{wen02} While the mean-field decoupling breaks the local $SU(2)\times SU(2)\times \cdots$ gauge symmetry, the gauge condition
\be
U_\rrs=W_\rr^\dagger U_\rrs W_\rs\label{igg}
\ee
is still fulfilled for a subset of transformations $W_\rr$. This can be seen by realizing that Eq.~(\ref{igg}) is {\it at least} satisfied for (site-independent) transformations of $\mathds{Z}_2$-type with $W_\rr \equiv W=+\mathds{1}_{4\times4}$ or $W_\rr \equiv W=-\mathds{1}_{4\times4}$. The subgroup of invariant transformations determines the IGG and the type of gauge fluctuations in an effective gauge theory. In the minimal case of a $\mathds{Z}_2$ IGG, the gauge-field excitations -- the so-called visons -- are gapped\cite{senthil00} and constitute an additional type of deconfined quasiparticle in a spin liquid (see below for details). In the following, we will restrict ourselves to $\mathds{Z}_2$ spin liquids since they are closest to a bare mean-field picture but still include long-range many-body entanglement with all its non-trivial implications for topological order and non-local excitations.

\subsection{Projective implementation of symmetries and PSG classification}
In this work, we investigate ``symmetric'' $\mathds{Z}_2$ spin liquids which do not spontaneously break lattice symmetries or time reversal invariance $\mathcal{T}$. For a square-lattice system in the $x$-$y$-plane with $\rr=(x,y)$ and $x,y=0,\pm1,\pm2,\ldots$ this means that the spin liquids need to be invariant under translations $T_x$, $T_y$ along both lattice directions [$T_x(\rr)=(x+1,y)$, $T_y(\rr)=(x,y+1)$], refections $P_x$, $P_y$ about the $x$ and $y$ axis [$P_x(\rr)=(-x,y)$, $P_y(\rr)=(x,-y)$] and a reflection $P_{xy}$ about the lattice diagonal [$P_{xy}(\rr)=(y,x)$]. Additionally, reflection symmetry about the lattice plane $P_z: z\rightarrow-z$ needs to be taken into account when SU(2) spin rotation symmetry is maximally lifted.\cite{reuther14} While this symmetry does not transform the site positions, it still has an effect in spin space since $z\rightarrow -z$ is a subgroup of SU(2) spin rotations. 

In the presence of a gauge freedom, a symmetry transformation $\mathcal{S}$ acts in two different ways (where $\mathcal{S}$ can be any of the above symmetries). Ignoring the gauge freedom, a symmetry transformation $\mathcal{S}$ first modifies $U_{\rrs}$ according to $U_{\rrs}\rightarrow U_{\mathcal{S}(\rr)\mathcal{S}(\rs)}$. The additional effect of the gauge invariance means that a symmetry operation $\mathcal{S}$ may always be supplemented with a gauge transformation $W_\rr$ leading to the {\it projective} implementation of symmetries $U_{\rrs}\rightarrow W_{\mathcal{S}(\rr)}^\dagger U_{\mathcal{S}(\rr)\mathcal{S}(\rs)}W_{\mathcal{S}(\rs)}$. As a consequence, a mean-field Hamiltonian $H_\text{mf}$ satisfies a symmetry $\mathcal{S}$ under the {\it weaker} condition that there exists a (site-dependent) gauge transformation $W^\mathcal{S}_\rr$ such that
\be
U_{\rrs}= W_{\mathcal{S}(\rr)}^{\mathcal{S}\dagger} U_{\mathcal{S}(\rr)\mathcal{S}(\rs)}W_{\mathcal{S}(\rs)}^\mathcal{S}\;.\label{psg_condition}
\ee
Comparing this relation with Eq.~(\ref{igg}), on sees that the elements of the IGG can be interpreted as the gauge transformation associated with the identity operation. For a given IGG, a PSG analysis classifies all possible projective implementations of symmetries -- characterized by the gauge transformations $W^\mathcal{S}_\rr$ -- that fulfill Eq.~(\ref{psg_condition}).

A PSG classification relies on the fact that the symmetries $\mathcal{S}$ (including the action of the corresponding gauge transformations $W^\mathcal{S}_\rr$) fulfill certain relations among each other. For example, two symmetry operations $\mathcal{S}_a$, $\mathcal{S}_b$ may commute, i.e., $\mathcal{O}_{ab}\equiv\mathcal{S}_a^{-1}\mathcal{S}_b^{-1}\mathcal{S}_a\mathcal{S}_b=\mathcal{I}$ (this is the case, e.g., for $\mathcal{S}_a=P_x$, $\mathcal{S}_b=P_y$), where $\mathcal{I}$ is the identity transformation. In a projective implementation, each individual operation in $\mathcal{O}_{ab}$ comes along with a gauge transformation. The total gauge transformation $W_\rr^{\mathcal{O}_{ab}}$ associated with $\mathcal{O}_{ab}$ is given by
\be
W^{\mathcal{O}_{ab}}_\rr=\left(W^{\mathcal{S}_a}_{\mathcal{S}^{-1}_b \mathcal{S}_a \mathcal{S}_b (\rr)}\right)^\dagger \left(W^{\mathcal{S}_b}_{\mathcal{S}_a \mathcal{S}_b (\rr)}\right)^\dagger W^{\mathcal{S}_a}_{\mathcal{S}_a \mathcal{S}_b (\rr)} W^{\mathcal{S}_b}_{\mathcal{S}_b (\rr)}\;.\label{w_example}
\ee
Since $W_\rr^{\mathcal{O}_{ab}}$ is the gauge transformation corresponding to the identity operation $\mathcal{I}$, it must be an element of the IGG. Therefore, it follows that either $W_\rr^{\mathcal{O}_{ab}}=+\mathds{1}_{4\times4}$ or $W_\rr^{\mathcal{O}_{ab}}=-\mathds{1}_{4\times4}$ on all sites. More generally, each sequence of symmetry transformations that yields the identity operation leads to two possibilities for choosing the sign of the associated total gauge transformation. Altogether, these signs characterize the different projective implementations of symmetries. The precise form of the gauge transformations follows from the $2\times2$ block structure of $W_\rr^\mathcal{S}$ consisting of matrices $w_\rr^\mathcal{S}$ [see Eq.~(\ref{gauge_trafo2})]. One finds that there is always a gauge in which these matrices are given by\cite{reuther14}
\be
w_\rr^\mathcal{S}=\eta_\rr^\mathcal{S} g_\mathcal{S}\;,
\ee
where $\eta^\mathcal{S}_\rr=\pm1$ is a site-dependent function and $g_\mathcal{S}$ is a spatially constant $2\times2$ SU(2) matrix. For the square lattice we will use the convenient gauge in which the $\eta^\mathcal{S}_\rr$ functions have the simple structure
\begin{eqnarray}
&\eta^\T_\rr=\eta_{\T}^{x+y}\;,\quad \eta^{T_x}_\rr=\eta^y\;,\quad \eta^{T_y}_\rr=1\;,\quad\eta_\rr^{P_z}=\eta_z^{x+y}&\notag\\
&\eta^{P_x}_\rr=\eta^x_1\eta^y_2\;,\quad \eta^{P_y}_\rr=\eta^x_2\eta^y_1\;,\quad\eta^{P_{xy}}_\rr=\eta^{xy}&\label{eta}
\end{eqnarray}
with $\eta_\mathcal{T}=\pm1$, $\eta=\pm1$, $\eta_z=\pm1$, $\eta_1=\pm1$, $\eta_2=\pm1$ independent of each other. Relations of the form of Eq.~(\ref{w_example}) can then be rewritten in terms of the $g_\mathcal{S}$ matrices
\be
g_{\mathcal{S}_a}^{-1}g_{\mathcal{S}_b}^{-1}g_{\mathcal{S}_a}g_{\mathcal{S}_b}=\pm\sigma^0\,.
\ee
Depending on the sign in this equation, the solutions $g_\mathcal{S}$ (if they exist) are either given by the identity $\sigma^0$ or by Pauli matrices $i\sigma^j$ (with $j=1,2,3$).

In total, for a given set of symmetries $\mathcal{S}$, the spatial sign pattern of $\eta^\mathcal{S}_\rr$ and the matrices $g_\mathcal{S}$ characterize a PSG and determine how projective symmetries act. When $\eta^\mathcal{S}_\rr=1$ on all sites and $g_\mathcal{S}=\sigma^0$, the projective version of $\mathcal{S}$ coincides with the ``naive'' implementation of the symmetry (i.e., in the absence of a gauge freedom). An exception is time-reversal $\mathcal{T}$, which we define such that $g^\mathcal{T}=i\sigma^2$ corresponds to the common implementation in a system with spinful fermions (in this implementation, an arbitrary single-particle wave function $|\phi\rangle$ acquires a minus sign under $\mathcal{T}^2$, i.e., $\mathcal{T}^2|\phi\rangle=-|\phi\rangle$). On the other hand, $g^\mathcal{T}=\sigma^0$ characterizes a system where time-reversal squares to one, $\mathcal{T}^2=1$, as is the case for spinless fermions.

When all gauge transformations $W^\mathcal{S}_\rr$ are known, Eq.~(\ref{psg_condition}) further puts constraints on the mean-field amplitudes $U_\rrs$. The precise form of the constraints in the channels $u_\rrs^s$, $u_\rrs^{t_1}$, $u_\rrs^{t_2}$, $u_\rrs^{t_3}$ is given in Appendix~\ref{app_a}. It is important to emphasize that these equations do not completely specify all parameters contained in $U_\rrs$, but rather relate amplitudes $U_\rrs$ and $U_{\mathcal{S}(\rr)\mathcal{S}(\rs)}$ with each other. Hence, a subset of all $U_\rrs$ (e.g. those for which $\drr\equiv(\dx,\dy)=\rs-\rr$ fulfills $\dx,\dy\geq0$ and $\dy\leq\dx$) serves as free parameters of a mean-field solution. Diagonalizing $H_\text{mf}$ finally yields the spinon-band structures in each projective representation as a function of these parameters.

A full classification of PSG representations for $\mathds{Z}_2$ spin liquids on the square lattice when SU(2) spin rotation symmetry is maximally lifted has previously been carried out in Ref.~\onlinecite{reuther14}. In the general case where the hopping and pairing mean-field amplitudes can be infinitely long-ranged, such an analysis yields 1760 different representations. This number also contains 272 SU(2) spin-rotation invariant states with $u_\rrs^{t_2}=u_\rrs^{t_3}\equiv0$ that have already been determined in the original work by X.-G. Wen.\cite{wen02} The remaining $1760-272=1488$ new representations are those in which the SU(2) symmetry is explicitly broken and the inversion symmetry $P_z$ acts non-trivially. In particular, it is shown in Ref.~\onlinecite{reuther14} that the finite $u_\rrs^{t_2}$ and $u_\rrs^{t_3}$ terms in these states have a form that generally admits chiral $p_x\pm ip_y$ pairing of the spinons. Depending on the particular PSG (e.g., on the implementation of time reversal) and the precise choice of the free mean-field parameters this may lead to a spin liquid with a non-trivial spinon-band topology {\it in addition} to the fractional and long-range entangled nature of the spinons. In the following Section~\ref{short-range} we will study these spin phases in the more realistic situation where the mean-field amplitudes are short-range (i.e., of nearest neighbor or second neighbor type only) which reduces the number of states enormously. Particularly, we will investigate the spinon band structures in a few cases and determine their topological properties. Thereafter, Section~\ref{visons} focuses on the properties of vison excitations and their coupling to spinons.

%====================================================================
\section{Short-range couplings and topological spinon bands}
\label{short-range}
\subsection{Nearest neighbor mean-field amplitudes}\label{nearest}
We first confine the range of the mean-field parameters $U_\rrs$ to nearest neighbors on the square lattice, where $\drr=\rs-\rr=(\pm1,0)$ or $\drr=(0,\pm1)$. This reduces the number of spin-liquid states drastically, because on the level of nearest neighbors many solutions vanish identically. Furthermore, even if solutions are finite, they might no longer have a $\mathds{Z}_2$ gauge structure, which means that Eq.~(\ref{igg}) is fulfilled for a set of transformations larger than $W_\rr=\pm1$ [such as U(1) transformations]. It is therefore crucial to check the IGG of the PSG mean-field solutions $U_\rrs$. We first briefly outline our approach to determine the IGG\cite{wen91,wen02} and then discuss the spinon properties of such solutions. 

The defining condition of the IGG [see Eq.~(\ref{igg})] formulates a relation between $W_{\rr_1}$ and $W_{\rr_2}$ for two nearest neighbor sites $\rr_1$, $\rr_2$,
\be
W_{\rr_1}=U_{\rr_1\rr_2}W_{\rr_2}U^{-1}_{\rr_1\rr_2}\;.
\ee
Inserting the analogous relation $W_{\rr_2}=U_{\rr_2\rr_3}W_{\rr_3}U^{-1}_{\rr_2\rr_3}$ yields
\be
W_{\rr_1}=U_{\rr_1\rr_2}U_{\rr_2\rr_3}W_{\rr_3}\left(U_{\rr_1\rr_2}U_{\rr_2\rr_3}\right)^{-1}\;.
\ee
Repeating this scheme for a sequence of nearest neighbor sites $\rr_1,\rr_2,\ldots,\rr_{n-1},\rr_n,\rr_1$ forming a closed loop $\mathcal{C}$, one obtains a condition for a {\it single} gauge operator $W_{\rr_1}$,
\be
W_{\rr_1}=P_\mathcal{C} W_{\rr_1}P_\mathcal{C}^{-1}\;,
\ee
where $P_\mathcal{C}=U_{\rr_1\rr_2}U_{\rr_2\rr_3}\cdots U_{\rr_{n-1}\rr_n}U_{\rr_n\rr_1}$. Writing $P_\mathcal{C}$ in $2\times2$ block form
\be
P_\mathcal{C}=
\left(\begin{array}{cc}
p^\mathcal{C}_{11} & p^\mathcal{C}_{12} \\
p^\mathcal{C}_{21} & p^\mathcal{C}_{22}
\end{array}\right)\;,
\ee
and using Eq.~(\ref{gauge_trafo2}) leads to the conditions
\be
[w_{\rr_1},p^\mathcal{C}_\kappa]=0\label{commutator}
\ee
which hold for all blocks $\kappa=11,12,21,22$ and loops $\mathcal{C}$, simultaneously. If, altogether, these constraints restrict $w_{\rr_1}$ such that the only possible solution is $w_{\rr_1}=\pm\sigma^0$, the IGG is proven to be $\mathds{Z}_2$. To evaluate the commutators, $p^\mathcal{C}_\kappa$ is expanded in terms of Pauli matrices and the identity matrix $p^\mathcal{C}_\kappa=\sum_{j=0}^3\alpha^{j\mathcal{C}}_{\kappa} \sigma^j$, and likewise for the gauge operation, $w_{\rr_1}=\alpha_{\rr_1}^0\sigma^0+i\sum_{j=1}^3\alpha_{\rr_1}^j\sigma^j$ (note that in the last equation, unitarity of $w_{\rr_1}$ requires the normalization of coefficients, $\sum_{j=0}^3|\alpha_{\rr_1}^j|^2=1$). The directional components of $p^\mathcal{C}_\kappa$ form vectors $(\alpha^{1\mathcal{C}}_{\kappa},\alpha^{2\mathcal{C}}_{\kappa},\alpha^{3\mathcal{C}}_{\kappa})$ in the three-dimensional coordinate space $\mathds{R}^3$. Calculating such vectors for all blocks $\kappa$ and loops $\mathcal{C}$ it is straightforward to show that if they form a {\it non-coplanar} set, Eq.~(\ref{commutator}) can only be fulfilled for $w_{\rr_1}=\pm\sigma^0$, which proves the $\mathds{Z}_2$ gauge structure. Otherwise, if these vectors span a plane, there is still a continuous set of $U(1)$ gauge transformations that fulfills Eq.~(\ref{commutator}).
   
Applying such an analysis to the aforementioned 1488 PSG solutions, we find that only 272 mean-field ans\"atze have finite nearest neighbor amplitudes. Further eliminating those with an IGG larger than $\mathds{Z}_2$ we finally identify 28 spin liquid phases which are characterized by the signs $\eta_\mathcal{T}=\pm1$, $\eta=\pm1$, $\eta_z=\pm1$, $\eta_1=\pm1$, $\eta_2=\pm1$ and the matrices $g_\mathcal{S}$ listed in Appendix~\ref{app_b}. Most strikingly, due to a subtle conflict of the effects of $P_z$, $\mathcal{T}$ and the requirement of a $\mathds{Z}_2$ gauge structure, all these solutions are characterized by $g^\mathcal{T}=\sigma^0$. This implies that time reversal squares to one, $\mathcal{T}^2=1$, and Kramer's degeneracy does not exists. As a consequence, there is no symmetry protection of boundary modes and the spinon bands are topologically trivial.\cite{fu07,fu07_2,hasan10} In other words, the systems belong to the class BDI\cite{altland97,hasan10} in which no topological index is defined (this is in contrast to systems with $g^\mathcal{T}=i\sigma^2$ which fall into the class DIII). Non-trivial spinon band structures on the square lattice and a $\mathds{Z}_2$ gauge structure can, therefore, only exist for mean-field amplitudes of (at least) second-neighbor range, as studied in Section~\ref{sec_neighbors}.

Even in the nearest neighbor case, the remaining 28 PSG solutions often have complicated spinon band structures which still strongly depend on the choice of a certain number of free mean-field parameters (there are typically three or four such parameters for each of the 28 non-vanishing nearest neighbor PSG solutions). By varying these amplitudes the systems may undergo transitions between phases with fully gapped band structures and those with discrete Dirac points in momentum space, indicating that individual PSG representations can again be subdivided into different phases.\cite{essin13} We refrain from developing a complete picture of all different spinon band structures in the remaining 28 solutions but instead, as an example, discuss the simplest case that we could identify.

\begin{figure}
\includegraphics[scale=0.3]{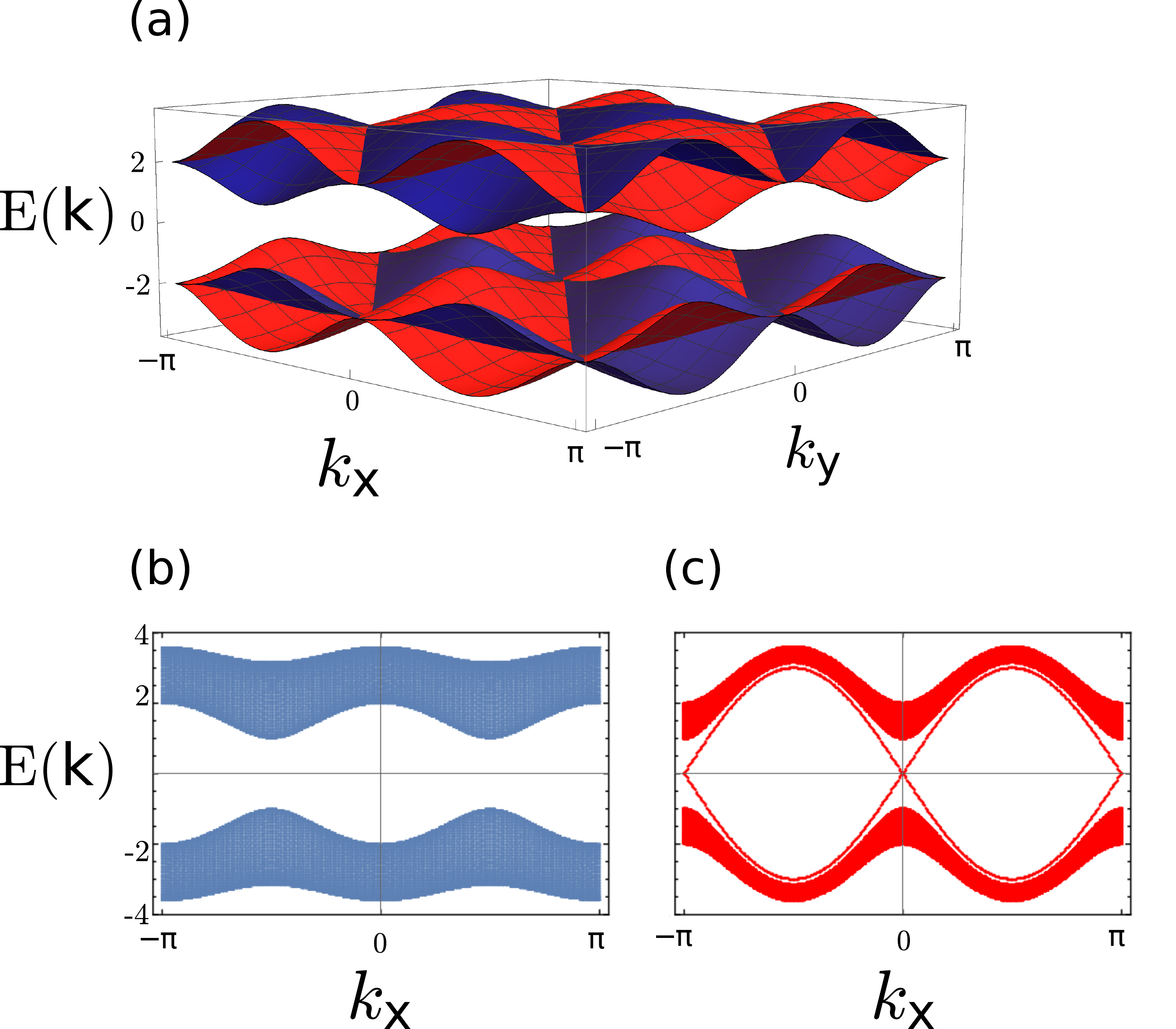}
\caption{Spinon-band structure of the nearest neighbor Hamiltonian in Eq.~(\ref{eq:sl01_k}) using the parameters $\alpha = \beta = \gamma =1$ and $\delta = 2$. The $h^1_\kk$ bands ($h^2_\kk$ bands) are plotted in red (blue) color. (a) Spinon bands for periodic boundary conditions in $x$ and $y$ directions. (b) Band structure of the $h^1_\kk$ block for a cylinder edge along the $x$ direction. (c) Band structure of the $h^2_\kk$ block for a cylinder edge along the $x$ direction. Note that for an edge along one of the lattice directions, only one block shows topologically trivial edge states.}
\label{fig:sl01_k}
\end{figure}

Interestingly, we found only one spin-liquid state where -- for a suitable choice of the spinor basis -- the matrix representation $U_\rrs$ has a block-diagonal form. The projective symmetry implementation of this state, characterized by the matrices $g_\mathcal{S}$ and the $\eta$ parameters in Eq.~(\ref{eta}), reads
\begin{eqnarray}
&g_{P_z}=i\sigma^3, \quad g_{\mathcal{T}}=\sigma^0, \quad g_{P_{xy}}=i\sigma^3,&\notag\\
&g_{P_x}=i\sigma^1, \quad g_{P_y}=i\sigma^1,&\notag\\
&\eta_{P_z}=1, \quad \eta_{\mathcal{T}}=-1, \quad \eta=1, \quad \eta_1=-1, \quad \eta_2=-1\;.&\notag\\
\end{eqnarray}
The block structure becomes obvious in the basis $\hat{\Psi}_\rr=(f_{\rr\uparrow},f_{\rr\uparrow}^\dagger,f_{\rr\downarrow},f_{\rr\downarrow}^\dagger)^\text{T}$ which groups together $\uparrow$ and $\downarrow$ parton operators. Transforming the mean-field Hamiltonian into $\kk$ space ($\hat{\Psi}_\rr\rightarrow \hat{\Psi}_\kk$) yields
\begin{equation}
H_\text{mf}=\sum_\kk\hat{\Psi}_\kk^\dagger  \begin{pmatrix}
h^1_\kk & 0\\ 0 & h^2_\kk
\end{pmatrix}\hat{\Psi}_\kk\label{eq:sl01_k}
\end{equation}
with
\begin{widetext}
\begin{align}
& h^1_\kk= \begin{pmatrix}
 (\alpha + \beta)\cos k_x + (\alpha - \beta)\cos k_y  & -(\gamma -\delta)\sin k_x +i(\gamma +\delta)\sin k_y\\ 
 (-\gamma +\delta)\sin k_x +i(\gamma +\delta)\sin k_y &  -(\alpha + \beta)\cos k_x - (\alpha - \beta)\cos k_y
\end{pmatrix}\notag \\
& h^2_\kk= \begin{pmatrix}
 (\alpha - \beta)\cos k_x + (\alpha + \beta)\cos k_y  & -(\gamma +\delta)\sin k_x +i(\gamma -\delta)\sin k_y\\ 
 -(\gamma +\delta)\sin k_x +i(\gamma -\delta)\sin k_y &  (-\alpha + \beta)\cos k_x - (\alpha + \beta)\cos k_y
\end{pmatrix}\;,
\end{align}
\end{widetext}
where $\alpha$, $\beta$, $\gamma$, $\delta$ are free (and real) parameters. Note that the block form of Eq.~(\ref{eq:sl01_k}) does not correspond to a U(1) spin-rotation symmetry around the $z$-axis (this would be the case for a block diagonal Hamiltonian in the original $\Psi_\rr$ basis). Rather, Eq.~(\ref{eq:sl01_k}) implies an invariance under a combined spin and particle-hole transformation. Due to $g^\mathcal{T}=\sigma^0$, time reversal does not transform the two blocks into each other and there is no simple relation between $h^1_\kk$ and $h^2_\kk$. If $\alpha$, $\beta$, $\gamma$, $\delta$ are all finite and $\gamma \neq \delta$ the spinon bands of Eq.~(\ref{eq:sl01_k}) are fully gapped and non-degenerate, as illustrated in Fig.~\ref{fig:sl01_k}. Even though the system is in a trivial phase we still find (topologically unprotected) boundary modes inside the bulk gap [see Fig.~\ref{fig:sl01_k}(b)]. Such states may generally appear in the vicinity of lattice inhomogeneities and their topologically trivial nature manifests in the fact that they are separated from the continuum of bulk states. Interestingly, for cylinder edges along the $x$-direction ($y$-direction) we only observe edge states in the $h^2_\kk$ block ($h^1_\kk$ block) but not in the $h^1_\kk$ block ($h^2_\kk$ block). This is again a consequence of the fact that the two blocks are not time-reversal related. We will revisit this spin state in Secion~\ref{visons} when we study the effects of vison excitations.  

\subsection{Second neighbor couplings and topological spinon bands}\label{sec_neighbors}
As outlined in the last section, topological spinon-bands cannot exist on the level of nearest neighbor models. To investigate systems with non-trivial bands we continue adding second neighbor mean-field amplitudes. As a result of the extra diagonal bonds, new types of loops can be formed such that the $\mathds{Z}_2$ gauge requirement is typically fulfilled more easily. In the first place, such an extension again drastically increases the number of states as compared to the nearest neighbor case. To keep the analysis manageable and to identify the simplest of such states, we impose certain constraints on the model parameters. Firstly, we only consider systems with $g^\mathcal{T}=i\sigma^2$ where Kramer's degeneracy allows for topologically projected edge modes. Secondly, the second neighbor mean-field parameters are assumed to be SU(2) spin-rotation invariant, i.e., of $u^s_\rrs$ type. In other words, we restrict ourselves to models where spin-orbit coupling only takes place on nearest neighbor bonds. This can be motivated by the fact that, taken individually, the effects of spin-orbit coupling and longer-ranged interactions are often sub-leading in real materials, such that the combination of both is expected to be even less important. Finally, to facilitate the analysis of topological invariants, we restrict ourselves to models with a simple block structure such as Eq.~(\ref{eq:sl01_k}). Under these assumptions, we find that there are only two different types of mean-field solutions. For special choices of the free parameters, some of their properties have already been discussed in Ref.~\onlinecite{reuther14}. In the following, we study these states in more detail (including gauge excitations) and map out their complete phase diagrams.

\subsubsection{First solution: BHZ-like model}
The first model is characterized by the projective symmetries
\begin{eqnarray}\label{eq:psgbhz}
&\mathit{g}_{P_z}=i\sigma^3, \quad \mathit{g}_{\mathcal{T}}=i\sigma^2, \quad \mathit{g}_{P_{xy}}=\sigma^0,&\notag\\
 &\mathit{g}_{P_x}=\sigma^0, \quad \mathit{g}_{P_y}=\sigma^0,&\notag\\
&\eta_{P_z}=1, \quad \eta_{\mathcal{T}}=1, \quad \eta= 1, \quad \eta_1=1, \quad \eta_2=1,&
\end{eqnarray}
leading to a Hamiltonian with three real constants $\alpha$, $\beta$, $\gamma$,
\begin{equation}
H_\text{mf}=\sum_\kk\hat{\Psi}_\kk^\dagger  \begin{pmatrix}
h_\kk & 0\\ 0 & h^*_{-\kk}
\end{pmatrix}\hat{\Psi}_\kk\label{eq:sl_bhz}
\end{equation}
where
\begin{widetext}
\begin{align}
h_{\kk}=& \begin{pmatrix}
\alpha\left( \cos k_x + \cos k_y \right) +  \beta \cos k_x \cos k_y &  \gamma \left( i\sin k_x - \sin k_y \right) \\
\gamma \left( i\sin k_x + \sin k_y \right) & -\alpha\left( \cos k_x + \cos k_y \right) - \beta\cos k_x \cos k_y
\end{pmatrix}\;.\label{eq:sl_bhz2}
\end{align}
\end{widetext}
Due to $g^\mathcal{T}=i\sigma^2$, the two blocks $h_\kk$ and $h^*_{-\kk}$ are time-reversal partners of each other. Most importantly, the terms $i\sin k_x \pm \sin k_y \sim i k_x\pm k_y$ induce the type of spin-momentum locking that generates non-trivial band structures. Indeed, Eq.~(\ref{eq:sl_bhz}) resembles the BHZ model\cite{bernevig06,konig08} which is a prototypical model for a topological insulator and has been used to describe the electronic bands of the quantum spin Hall material HgTe. The difference is that $h_\kk$ in Eq.~(\ref{eq:sl_bhz2}) exhibits a term $\sim\beta\cos k_x \cos k_y\sigma^3$ instead of $\sim [M-B(k_x^2+k_y^2)]\sigma^3$. Both terms induce a negative (positive) mass in the upper (lower) band around the $\Gamma$-point, as needed for a topological band structure. The momentum dependence of $\beta\cos k_x \cos k_y\sigma^3$ away from the $\Gamma$-point, however, also generates phases which are not present in the BHZ model, as discussed in the following.
\begin{figure*}
\includegraphics[width=0.8\linewidth]{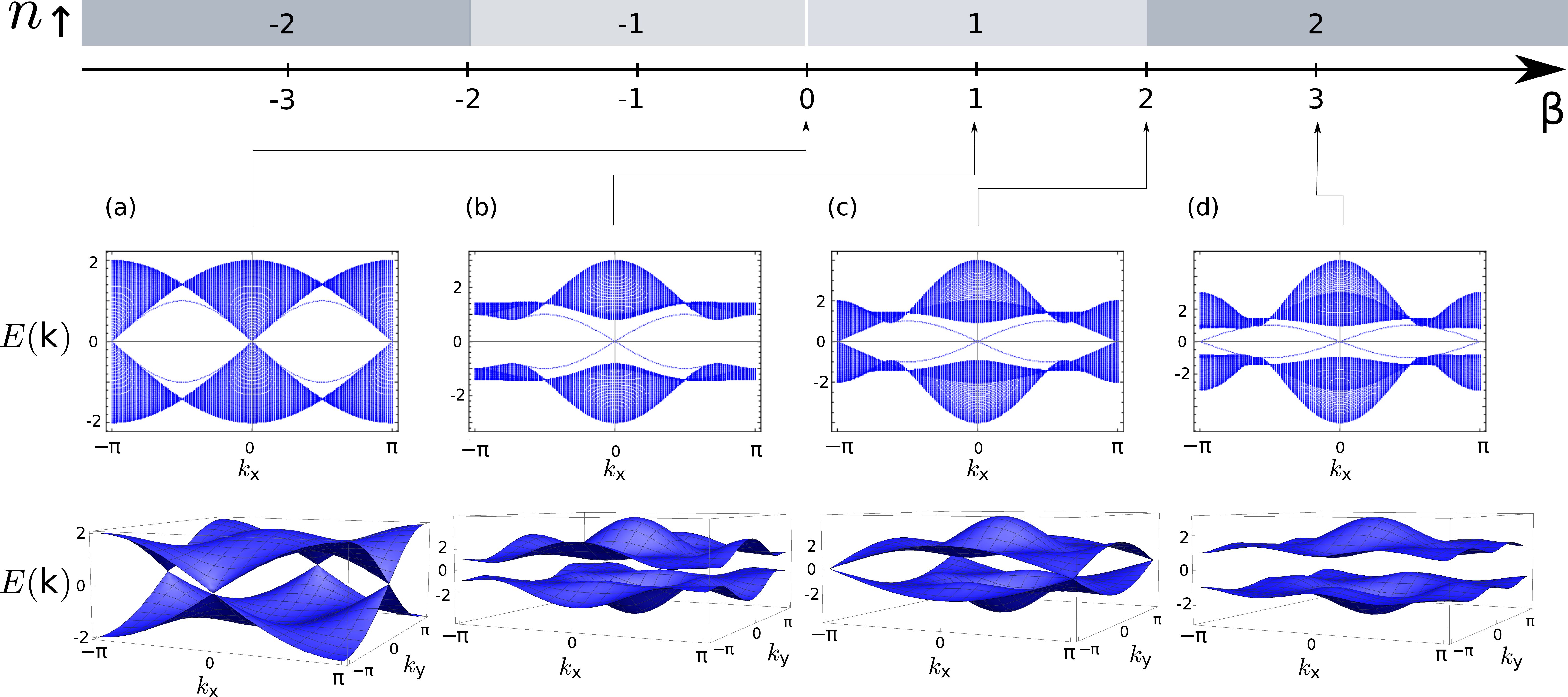}
\caption{Top: Phase diagram of the topological spinon bands for the mean-field Hamiltonian in Eq.~(\ref{eq:sl_bhz}) using $\alpha=1$. Chern numbers $n_\uparrow$ of the block $h_\kk$ are indicated in each phase. Note that the phase diagram is independent of $\gamma$. (a)-(d) Spinon band structures for $\beta=0,1,2,3$ and $\alpha=\gamma=1$ where the upper plots correspond to a cylinder geometry with an edge along the $x$-axis and the lower plots are for a torus geometry. All spinon bands are doubly degenerate with respect to the spin degree of freedom.}
\label{fig:bhz}
\end{figure*}

Setting $\alpha=1$ and varying $\beta$, the system goes through a sequence of different phases as illustrated in Fig.~\ref{fig:bhz}. Note that the parameter $\gamma$ sets the overall size of the topological gap but does not shift the phase boundaries. Since the two blocks $h_\kk$ and $h^*_{-\kk}$ with Chern numbers $n_\uparrow$ and $n_\downarrow$, respectively, are time-reversal partners, one finds $n_\uparrow=-n_\downarrow$ in each phase. Furthermore, $\beta\rightarrow-\beta$ reverses the signs of both Chern numbers but otherwise leaves the band topologies unchanged.  At $\beta=0$, the bulk has gapless nodes at $\kk=(0,\pm\pi)$ and $\kk=(\pm\pi,0)$, see Fig.~\ref{fig:bhz}(a). A finite term $\beta>0$ opens gaps at these points, leading to a topological phase with Chern numbers $n_\uparrow=1$, $n_\downarrow=-1$ and a pair of helical boundary modes crossing each other at $k_x=0$ [Fig.~\ref{fig:bhz}(b)]. In this phase the system features a non-trivial $\mathds{Z}_2$ topological invariant\cite{hasan10,roy09} given by $\nu\equiv\frac{n_\uparrow-n_\downarrow}{2}\text{mod}\,2=1$. While the phenomenology of the bands is in complete analogy to a topological insulator, it is worth emphasizing that the interpretation is rather different here. Since our quasiparticles are spinons and the ``sin''-terms in Eq.~(\ref{eq:sl_bhz2}) describe spinon pairing, the system can be considered as two copies of a topological ``spinon superconductor'' with opposite chiralities.\cite{qi09,qi10,qi11} Consequently, the counter-propagating edge states are Majorana zero modes $\gamma_\text{R}$, $\gamma_\text{L}$ and a mass term $i\gamma_\text{R}\gamma_\text{L}$ gapping out these states is forbidden by time-reversal symmetry.

Increasing $\beta$, the system undergoes another transition at $\beta=2$ where the bulk gap closes at $\kk=(\pm\pi,\pm\pi)$, as shown in Fig.~\ref{fig:bhz}(c). Above this point the gap reopens [Fig.~\ref{fig:bhz}(d)] and Chern numbers are given by $n_\uparrow=2$, $n_\downarrow=-2$ (note that this phase is not present in the BHZ model). Accordingly, the edge spectrum exhibits {\it two} pairs of counter-propagating Majorana modes crossing each other at $k_x=0$ and $k_x=\pi$. This regime extends up to $\beta\rightarrow\infty$ such that, in total, Eq.~(\ref{eq:sl_bhz}) never exhibits trivial bands with vanishing Chern numbers. This can be traced back to the fact that the terms $[\alpha\left( \cos k_x + \cos k_y \right) +  \beta \cos k_x \cos k_y]\sigma^3$ in Eq.~(\ref{eq:sl_bhz2}) always have a finite ``spinon Fermi-surface'' such that the additional $\gamma$ terms can open a topological gap at these surfaces. Due to the Chern numbers $n_{\uparrow/\downarrow}=\pm2$, the $\mathds{Z}_2$ topological invariant is trivial ($\nu=0$) and one would expect that the protection of boundary modes is lost. However, as already discussed in Ref.~\onlinecite{reuther14}, when taking into account lattice symmetries, there is still a protection of the edge states. Denoting the Majorana zero modes at $k_x=0$ ($k_x=\pi$) by $\gamma_\text{R}$ and $\gamma_\text{L}$ ($\eta_\text{R}$ and $\eta_\text{L}$) one finds that the mass terms $i\gamma_\text{R}\gamma_\text{L}$ and $i\eta_\text{R}\eta_\text{L}$ are forbidden due to time-reversal symmetry. On the other hand, the coupling terms $i\gamma_\text{R}\eta_\text{L}$ or $i\gamma_\text{R}\eta_\text{L}$ gapping out boundary states at {\it different} $k_x$ require a finite momentum transfer $\Delta k_x=\pi$ which is only possible when translation symmetry in $x$-direction is broken. (Note that in this PSG, terms $i\gamma_\text{R}\eta_\text{L}$ and $i\gamma_\text{R}\eta_\text{L}$ are also forbidden due to the $P_z$ symmetry.) Hence, as long as translation and time-reversal symmetries are intact, the edge modes must remain gapless. Since this protection relies on lattice symmetries, the system can be considered as a spinon version of a topological {\it crystalline} superconductor.\cite{fu11,slager13,hsieh12}

\subsubsection{Second solution: BHZ-like model with spatially dependent implementation of time reversal}
The second mean-field solution that satisfies the afore-stated conditions has the projective symmetry implementation
\begin{eqnarray}\label{eq:psgbhz}
&\mathit{g}_{P_z}=i\sigma^3, \quad \mathit{g}_{\mathcal{T}}=i\sigma^2, \quad \mathit{g}_{P_{xy}}=\sigma^0,&\notag\\
&\mathit{g}_{P_x}=i\sigma^3, \quad \mathit{g}_{P_y}=i\sigma^3,&\notag\\
&\eta_{P_z}=1, \quad \eta_{\mathcal{T}}=-1, \quad \eta= 1, \quad \eta_1=-1, \quad \eta_2=-1\;.&\notag\\
\end{eqnarray}
The corresponding Hamiltonian reads
\begin{equation}
H_\text{mf}=\sum_\kk\hat{\Psi}_\kk^\dagger  \begin{pmatrix}
h_\kk & 0\\ 0 & h^*_{-\kk+(\pi,\pi)}
\end{pmatrix}\hat{\Psi}_\kk\label{eq:sl_type2}
\end{equation}
with
\begin{widetext}
\begin{align}
h_{\kk}=& \begin{pmatrix}
\alpha\left( -\cos k_x + \cos k_y \right) +  \beta \cos k_x \cos k_y &  \gamma \left( i\sin k_x - \sin k_y \right) \\
\gamma \left( i\sin k_x + \sin k_y \right) & \alpha\left( \cos k_x - \cos k_y \right) - \beta\cos k_x \cos k_y
\end{pmatrix}\;,\label{eq:sl_type22}
\end{align}
\end{widetext} 
where $\alpha$, $\beta$, $\gamma$ are three real parameters. Interestingly, time reversal now has a non-trivial real-space structure given by $\eta_\rr^\mathcal{T}=(-1)^{x+y}$ which in momentum space corresponds to a shift $\kk\rightarrow \kk+(\pi,\pi)$. Apart from complex conjugation and $\kk\rightarrow-\kk$, the two time-reversal-related blocks in Eq.~(\ref{eq:sl_type2}) therefore also differ by a wave vector $(\pi,\pi)$. Compared to the previous BHZ-like model, here, all bands of the upper block (lower block) are shifted by $k_x\rightarrow k_x+\pi$ and $k_y$ unchanged ($k_y\rightarrow k_y+\pi$ and $k_x$ unchanged). On the other hand, the topological phase diagram and Chern numbers remain the same and will not be discussed again.

In summary, this analysis shows that for a spin-anisotropic $\mathds{Z}_2$ spin liquid on the square lattice, topological spinon bands are only possible for second neighbor coupling terms. If such terms are present, non-trivial band structures naturally appear even under the most simplifying assumptions.

%====================================================================
\section{Vison excitations}
\label{visons}
\subsection{Effective $\mathds{Z}_2$ gauge theory and static approximation}\label{review_gauge_theory}
The bare mean-field models studied so far need to be treated with caution since their eigenstates generically violate the parton constraint and therefore do not even represent proper spin states. This is equivalent to the observation that the mean-field Hamiltonians are not invariant under a general local SU(2) gauge transformation $W_\rr$. The problem is obviously rooted in the fact that we assumed the matrices $U_\rrs$ to be constant objects instead of fluctuating fields. To correct this deficiency and restore a gauge freedom, we consider the minimal set of fluctuations in $U_\rrs$ given by
\be
U_\rrs \rightarrow U_\rrs \sigma^z_\rrs\;,\label{gauge_fluctuation}
\ee
where $\sigma^z_\rrs=\pm1$ is a $\mathds{Z}_2$ gauge-field variable defined on the {\it bonds} of the lattice. Instead of Eq.~(\ref{ham_mf}) the model Hamiltonian then reads\cite{senthil00}
\be
H=\sum_\pair\left( \Psi_\rr^\dagger \sigma^z_\rrs U_\rrs\Psi_\rs+\text{H.c.}\right)\;.\label{ham_z2}
\ee
There are various reasons for choosing $\mathds{Z}_2$ gauge fields $\sigma_\rrs^z$. First, it is the simplest extension of a bare mean-field theory which still generates non-trivially correlated phases described by lattice gauge theories\cite{wegner71,kogut79,senthil00} (see below). Second, it can be shown that the fluctuations in $U_\rrs$ are dictated by the IGG of the corresponding PSG mean-field solution\cite{wen02} such that Eq.~(\ref{gauge_fluctuation}) is consistent with the systems studied in Section~\ref{short-range}. Finally, there exists a substantial number of spin systems -- the Kitaev honeycomb model\cite{kitaev06} being one of the most prominent ones -- where a theory of the form of Eqs.~(\ref{gauge_fluctuation}) and (\ref{ham_z2}) corresponds to an {\it exact} rewriting of the original spin Hamiltonian.\cite{baskaran07,yao07,feng07,knolle14,brien16} For strongly frustrated and magnetically disordered spin systems where such a rewriting does not exist, it is widely believed that Eq.~(\ref{ham_z2}) at least provides a good description of the low energy fractional degrees of freedom.
\begin{figure}
\includegraphics[width=0.6\linewidth]{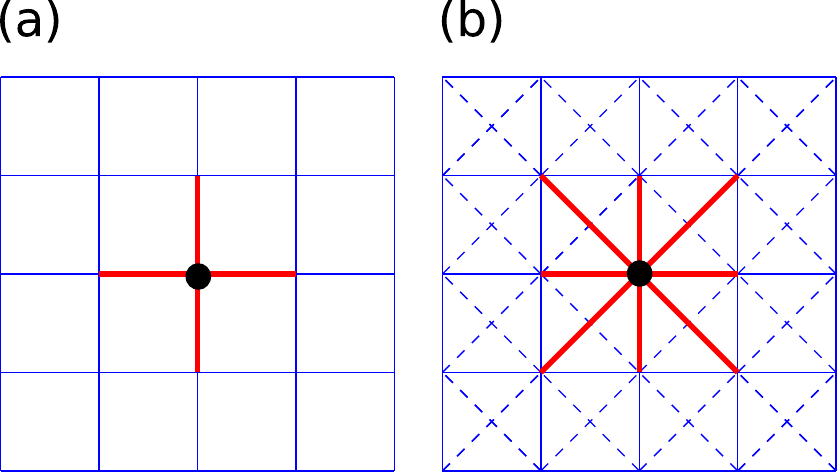}
\caption{Effect of a local gauge transformation $\mathcal{S}_\rr$ [Eq.~(\ref{star})] acting on a site $\rr$ (black dot): All bond variables $\sigma^z_\rrs$ are flipped ($\sigma^z_\rrs\rightarrow-\sigma^z_\rrs$) on links forming a star around $\rr$  (red lines). (a) For a nearest neighbor mean-field Hamiltonian the star consists of four bonds emanating from $\rr$. When second neighbor mean-field amplitudes are added (dashed lines), the gauge fields $\sigma^z_\rrs$ are also defined on diagonal links, such that the gauge transformation acts on all bonds highlighted in (b).}
\label{fig:star}
\end{figure}

Taking into account the gauge fluctuations, Eq.~(\ref{ham_z2}) satisfies a {\it local} $\mathds{Z}_2$ gauge invariance $\mathcal{G}$ given by the {\it combined} transformation
\be 
\mathcal{G}:\;\Psi_\rr\rightarrow -\Psi_\rr\;,\quad \sigma_\rrs^z\rightarrow -\sigma^z_\rrs\;,\label{gauge_trafo3}
\ee
where $\rs$ are all nearest and/or second neighbor sites of $\rr$ (depending on the range of the parameters $U_\rrs$). Interpreting the fields $\sigma^z_\rrs$ as Pauli matrices, the sign flips in Eq.~(\ref{gauge_trafo3}) can also be expressed in terms of $\sigma^x_\rrs$ operators, yielding
\be 
\mathcal{G}:\;\Psi_\rr\rightarrow -\Psi_\rr\;,\quad \sigma_\rrs^z\rightarrow \mathcal{S}_\rr^\dagger\sigma^z_\rrs\mathcal{S}_\rr\;,\label{gauge_trafo4}
\ee
with
\be
\mathcal{S}_\rr=\prod_{\rs\in\text{star}(\rr)}\sigma_\rrs^x\label{star}
\ee
and star$(\rr)$ denotes all (nearest and/or second) neighbor sites that form a star around $\rr$ (see Fig.~\ref{fig:star}). The existence of a gauge freedom again indicates that the physical Hilbert space is smaller than that of the $\Psi_\rr$ and $\sigma^z_\rrs$ degrees of freedom. The constraint selecting the physical states now takes the form
\be
\mathcal{S}_\rr(-1)^{f_{\rr\uparrow}^\dagger f_{\rr\uparrow}+f_{\rr\downarrow}^\dagger f_{\rr\downarrow}}=-1\label{constraint}
\ee
for all sites $\rr$.\cite{frustration_book,senthil00,senthil01}

When the spinons $\Psi_\rr$ are gapped, they can formally be integrated out yielding an effective low energy theory for the gauge fields. The resulting pure gauge theory can contain all types of gauge-invariant operators, i.e., those that commute with $\mathcal{S}_\rr$ on all sites $\rr$.\cite{wen_book,frustration_book,moessner01,misguich02,nikolic03} Terms that fulfill this condition are $\sigma^x_\rrs$ fields or {\it loops} of $\sigma^z_\rrs$ operators. Taking into account the lowest orders (i.e. only the shortest loops in $\sigma^z_\rrs$ and terms linear in $\sigma^x_\rrs$) yields the model
\begin{align}
H_\text{gauge}=&-h_1\hspace*{-2pt}\sum_{\langle\rrs\rangle}\sigma_\rrs^x-h_2\hspace*{-2pt}\sum_{\langle\langle\rrs\rangle\rangle}\sigma_\rrs^x-K_1\sum_{\Box}\prod_{\rr,\rs\in\Box}\hspace*{-2pt}\sigma_\rrs^z\notag\\
&-K_2\Bigg(\sum_{\drs}\prod_{\rr,\rs\in\drs} \sigma_\rrs^z+\sum_{\dls}\prod_{\rr,\rs\in\dls} \sigma_\rrs^z\notag\\
&+\sum_{\uls}\prod_{\rr,\rs\in\uls} \sigma_\rrs^z+\sum_{\urs}\prod_{\rr,\rs\in\urs} \sigma_\rrs^z\Bigg)-\ldots\;.\label{pure_gauge}
\end{align}
Here, $\langle\rrs\rangle$ ($\langle\langle\rrs\rangle\rangle$) denotes nearest (second) neighbor pairs of sites and $\Box$ stands for the unit squares of the lattice. The notation $\rr,\rs\in\Box$ means that the sites $\rr$, $\rs$ belong to one of the square edges. The same convention is used for pairs $\rr,\rs\in\drb,\dlb,\ulb,\urb$ where $\drb,\dlb,\ulb,\urb$ are the four types of elementary triangles with one diagonal link. Note that the terms $\sim h_2$ and $\sim K_2$ only appear for models with second neighbor mean-field amplitudes.  

\begin{figure*}
\includegraphics[width=0.99\linewidth]{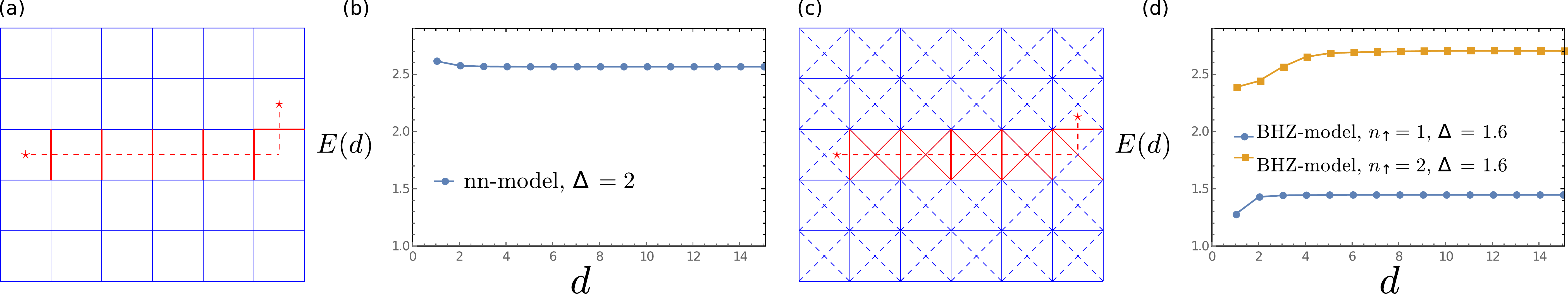}
\caption{(a) Possible gauge field configuration for a pair of separated visons (marked by stars) in the case of nearest neighbor mean-field models. The full red lines indicate bonds with $\sigma^z_\rrs=-1$ while $\sigma^z_\rrs=1$ otherwise. Gauge fields $\sigma^z_\rrs=-1$ occur on all bonds crossing the ``gauge string'' (dashed red line). The plaquettes at which the visons reside are threatened by a finite local flux. (b) Vison pair-excitation energy $E(d)$ as a function of the distance $d$ for the nearest neighbor PSG solution in Eq.~(\ref{eq:sl01_k}) using the parameters $\alpha = \beta = \gamma =1$, $\delta =2$. Here, $d$ is defined in units of the nearest-neighbor lattice spacing and the separation is along a lattice direction. For comparison, the spinon bulk gap $\Delta$ is indicated in the figure. (c) Same as (a) but with additional second neighbor mean-field amplitudes. In this case also diagonal bond variables $\sigma^z_\rrs$ need to be flipped along the gauge string. (d) Vison potentials $E(d)$ for the BHZ-like model [see Eq.~(\ref{eq:sl_bhz})] in the phases with Chern numbers $n_\uparrow=1$ (blue line with parameters $\alpha=\beta=\gamma=1$) and $n_\uparrow=2$ (yellow line with parameters $\alpha=\gamma=1$, $\beta=3$).}
\label{fig:vison_separation}
\end{figure*}

The pure two-dimensional gauge theory in Eq.~(\ref{pure_gauge}) is known to have two different phases: A confined and a deconfined phase.\cite{savary16,wen_book,frustration_book} When the $K$-terms (also referred to as ``magnetic'' terms or fluxes) are much larger than the $h$-terms (often called ``electric'' fields), the ground state is given by the configuration where all square-loop (and triangular-loop) operators fulfill $\sigma^z_{\rr_1\rr_2}\sigma^z_{\rr_2\rr_3}\sigma^z_{\rr_3\rr_4}\sigma^z_{\rr_4\rr_1}=1$. Consequently, excitations correspond to configurations with negative loops terms, $\sigma^z_{\rr_1\rr_2}\sigma^z_{\rr_2\rr_3}\sigma^z_{\rr_3\rr_4}\sigma^z_{\rr_4\rr_1}=-1$, each associated with an excitation energy $\sim K$. Together with the spinons, these fluxes (also called visons) represent the two types of fundamental quasiparticles in a spin liquid. In similarity to the spinons, the visons can only be created in pairs. Most importantly, if the electric fields $\sim h$ are sufficiently small, there is no long-range binding force between the visons such that they are effectively free, i.e., deconfined. It is important to emphasize that the absence of visons does not necessarily mean that $\sigma^z_\rrs=1$ on all bonds. Indeed, one can easily see that on a torus there are four {\it gauge inequivalent} ground-state configurations without any visons [they correspond to the configurations where the gauge strings illustrated in Fig.~\ref{fig:vison_separation}(a), (c) wind around non-contractible loops of a torus]. This ground-state degeneracy is topologically protected as it cannot be lifted without closing the vison gap. A gauge theory in this phase is relevant for a low-energy description of $\mathds{Z}_2$ quantum spin liquids, since it correctly captures their long-range entangled and topological properties.

In the limit where the $h$-terms are dominant, the system can (in lowest order) be mapped onto a model of non-interacting Ising spins in a magnetic field. In this topologically trivial phase the non-degenerate ground state is given by the configuration where $\sigma^x_\rrs=1$ on all bonds and the visons experience a long-range confining force.

Coming back to Eq.~(\ref{ham_z2}), which is the starting point of the analysis in the next section, the gauge fields only appear through one type of Pauli matrix $\sigma^z_\rrs$. On the level of the Hamiltonian it therefore seems that the fields $\sigma^z_\rrs$ are static Ising variables. The gauge fields, however, become dynamic if one takes into account the constraint in Eq.~(\ref{constraint}) since a {\it single} bond variable $\sigma_\rrs^z$ does not commute with $\mathcal{S}_\rr$. With this constraint, Eq.~(\ref{ham_z2}) becomes a complicated many-body problem that -- in similarity to the original spin model -- cannot be easily solved. The most straightforward approximation that allows us to proceed, is to treat Eq.~(\ref{ham_z2}) as it is, but simply ignore the gauge constraint in Eq.~(\ref{constraint}). While this might first appear as a very crude simplification, it is conceptually similar to the PSG approach where, likewise, spinon band structures are determined without taking into account the parton constraint. Ignoring Eq.~(\ref{constraint}) means that the gauge fields become static and the pure gauge theory in Eq.~(\ref{pure_gauge}) does not contain any $\sigma^x_\rrs$ terms. In the case of a $\mathds{Z}_2$ spin liquid this can change details of the low energy properties; for example, vison dispersions become flat. On the other hand, the key properties of $\mathds{Z}_2$ gauge theories such as vison deconfinement, finite vison gaps, and topological ground-state degeneracies are independent of the gauge field dynamics. We note that the situation is similar to Kitaev spin models on tri-coordinated lattices,\cite{kitaev06} where the gauge fields are likewise found to be static. In that sense, a static approximation can be considered as a convenient way of studying generic quasiparticle properties of $\mathds{Z}_2$ spin liquids, without the need to solve a complicated many-body problem. 

\subsection{Vison deconfinement in selected PSG solutions}\label{deconfinement}
Using the static approximation discussed in the last section, the spinon and vison degrees of freedom in Eq.~(\ref{ham_z2}) can be treated separately, i.e, for each fixed configuration of the gauge fields $\sigma_\rrs^z$, a free fermionic model in the spinons needs to be solved. This procedure is well-known from Kitaev spin models but, to the best of our knowledge, has not been systematically applied to PSG solutions. While the ``electric'' fields $\sim h$ are not accessible within a static scheme, one may still estimate the vison masses $\sim K$ and confirm that visons are indeed deconfined.

To calculate vison creation and separation energies, we introduce a pair of fluxes by changing the signs of $\sigma^z_\rrs$ on all bonds crossing a line between the vison cores, see Fig.~\ref{fig:vison_separation}(a), (c). Note that for second neighbor mean-field terms it is important to also flip the signs of the gauge fields on diagonal bonds along the string, as shown in Fig.~\ref{fig:vison_separation} (c) (otherwise, local fluxes $\sim K_2$ would be finite along the string, creating a {\it chain} of visons). The effective vison-pair potential $E(\boldsymbol{d})$ (where $\boldsymbol{d}$ is the vector between the vison cores) is obtained from the total energy of the two-vison state minus the ground-state energy $E_0$ of the flux-free state.  As an example, we illustrate $E(\boldsymbol{d})$ for the nearest neighbor PSG solution in Eq.~(\ref{eq:sl01_k}) and for the second neighbor BHZ-like model in Eq.~(\ref{eq:sl_bhz}), where for the latter system we distinguish between phases with Chern numbers $n_\uparrow=1$ and $n_\uparrow=2$, see Fig.~\ref{fig:vison_separation}. (Since the PSG solution in Eq.~(\ref{eq:sl_type2}) differs from the BHZ-like model in Eq.~(\ref{eq:sl_bhz}) only by shifts in momentum space, it does not have distinct properties and will not be further considered here.)

For all models that we have studied and independent of the Chern numbers we find that $E(d)\equiv E(\boldsymbol{d}=(d,0))$ (where visons are separated along a lattice direction) already saturates after a few lattice spacings, demonstrating that confining forces between the visons are of very short-range nature. In the case of the BHZ-like model, visons experience a mild attraction at small distances. Interestingly, the excitation energy $E(d=1)$ for a pair of nearest neighbor visons agrees with the asymptotic value $E(d\rightarrow\infty)$ within $\sim10\%$ or less. This indicates that in a pure gauge theory description [see Eq.~(\ref{pure_gauge})] obtained by integrating out the spinons, additional contributions with loops longer than the $K_1$ and $K_2$ terms must be small. The nearest neighbor energy $E(d=1)$ therefore provides a good estimate for the size of the lowest order $K$-terms. Comparing $E(d\rightarrow\infty)$ with the spinon bulk gap $\Delta$, we find that for all models studied, the vison mass is roughly on the order of $\Delta$ (the corresponding numbers are given in Fig.~\ref{fig:vison_separation}). We note that the deconfined property of visons is already expected from the structure of the gauge theory in Eq.~(\ref{ham_z2}). This is because, by successively applying gauge transformations $\mathcal{S}_\rr$, the gauge string between two visons can be arbitrarily deformed without moving the visons. The length of the string is, therefore, no physical observable and cannot induce confining.  
\begin{figure}
\includegraphics[width=0.99\linewidth]{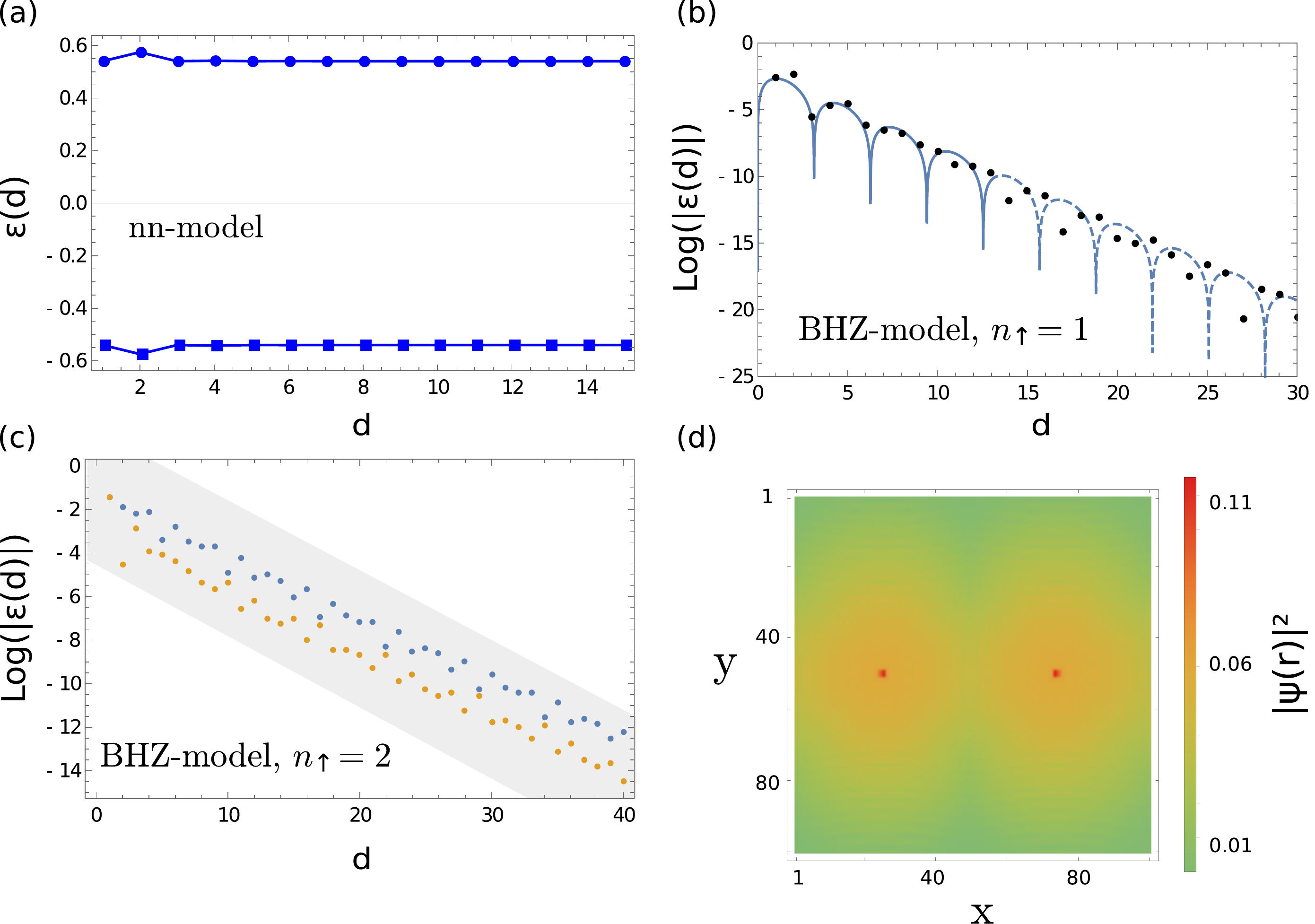}
\caption{(a)-(c) Excitation energies $\epsilon(d)$ of in-gap states as a function of the vison-pair separation $d$ for (a) the nearest neighbor spin liquid in Eq.~(\ref{eq:sl01_k}), (b) the BHZ-like spin liquid in the phase with $n_\uparrow=1$ [see Eq.~(\ref{eq:sl_bhz})] and (c) the BHZ-like spin liquid in the phase with $n_\uparrow=2$. The corresponding mean-field parameters are the same as in Fig.~\ref{fig:vison_separation}. The plots in (b) and (c) use a logarithmic energy axis and only show the positive part of the spectrum (the negative part is an exact mirror image). Note that all depicted data points are degenerate with respect to the spin degree of freedom. In (b) we find indications for an oscillating modulation as indicated by a fit of $\epsilon(d)$ to a function $\sim e^{-ad}\cos(bd+\phi)$ with fit parameters $a$, $b$, $\phi$ (blue line). The quality of the fit becomes worse at larger $d$, see dashed blue line. (d) Real-space probability distribution of the spinon wave function corresponding to the zero energy mode of the $n_\uparrow=1$ BHZ-like system in (b). The locations of the vison cores are given by the red spots where the wave function is sharply peaked.}
\label{fig:bound_modes}
\end{figure}

\subsection{Spinon-vison bound states}\label{bound_states}
Since visons represent point defects of the fermionic system, they modify the spinon spectrum, possibly leading to spinon-vison bound states inside the bulk gap. Here, we study spinon-vison bound states for the three models discussed in last section and monitor their energies $\epsilon(d)$ as a function of the vison separation $d$ (where $d$ again measures their distance along a lattice direction). Particularly, we connect the existence of zero-energy modes to the topology of the spinon bands (for a related work, see Ref.~\onlinecite{slager15}).

We first calculate the fermionic spinon spectrum for the nearest-neighbor model in Eq.~(\ref{eq:sl01_k}) in the presence of two visons. Since all models studied here contain superconducting spinon pairing terms, the spectrum is particle-hole symmetric and it is sufficient to consider the positive part of the spectrum only. For the nearest neighbor model we find two degenerate in-gap modes (one for each spin direction) at finite energies binding to each vison, see Fig.~\ref{fig:bound_modes}(a). With increasing vison distance, the energies $\epsilon(d)$ of these states quickly saturate and remain constant (and finite) for large $d$. As a generic example of a fermionic model without any topological invariant, bound states can always exist, however, they are not protected by a symmetry. Whether or not they appear depends on the details of the Hamiltonian. By changing the model parameters, the bound states can, in principle, be shifted into the continuum of bulk states without traversing a phase transition.

Bound states in the BHZ-like model of Eq.~(\ref{eq:sl_bhz}) show a distinctly different behavior. Considering the phase with Chern number $n_\uparrow=1$, a pair of visons binds two degenerate mid-gap fermionic spinon modes (one for each spin direction) as illustrated in Fig.~\ref{fig:bound_modes}(b). In contrast to the trivial band structure of the nearest neighbor model, the energy $\epsilon(d)$ of this state shows a rapid exponential decrease as a function of $d$. By mapping the visons onto vortices in a topological $p+ip$ superconductor we will show below that the bound states at $d\rightarrow\infty$ are indeed exactly described by two Majorana zero modes\cite{alicea12} $\gamma_\uparrow$ and $\gamma_\downarrow$ associated with each vison core. Therefore, the existence of zero modes is a topologically protected property that directly follows from the non-trivial Chern number of the bulk bands. In similarity to the one-dimensional edge states discussed above, a coupling term $i\gamma_\uparrow \gamma_\downarrow$, gapping out the zero modes, is forbidden due to time-reversal symmetry. The finite gaps at small $d$ are due to the spatial overlap of the Majorana wave functions localized at different vison cores. This is illustrated in Fig.~\ref{fig:bound_modes}(d), showing wave functions sharply peaked at the vison positions and exponentially decaying tails. The quasiparticle excitations within the bulk gap can, therefore, be considered as composite objects consisting of one vison and two Majorana modes.

While globally, the binding energies in Fig.~\ref{fig:bound_modes}(b) follow an exponential decrease, we also observe local deviations from this behavior. To a certain degree, this can be explained by a modulation of $\epsilon(d)$ with an oscillating function, even though not all data points follow this trend (particularly at larger distances). Interestingly, a similar observation has been reported for spin liquids on the Kitaev honeycomb model with a superlattice of visons, showing a nucleation of Majorana fermion bands with different Chern numbers.\cite{lahtinen12,lahtinen14} In these models the oscillating behavior has been attributed to the fusion rules of Ising anyons.

We finally consider bound states of the BHZ-like model in the phase with Chern number $n_\uparrow=2$. The in-gap spectrum effectively corresponds to two copies of the bound states of the $n_\uparrow=1$ case, i.e., we find two fermionic modes with exponentially decaying energies for each spin sector, see Fig.~\ref{fig:bound_modes}(c). In terms of Majorana degrees of freedom there are now four zero modes $\gamma_{\uparrow}$, $\eta_{\uparrow}$, $\gamma_{\downarrow}$, $\eta_{\downarrow}$ tied to each vison core. In similarity to the $n_\uparrow=1$ phase, a finite gap generated by coupling terms $i\gamma_{\uparrow}\gamma_{\downarrow}$ or $i\eta_{\uparrow}\eta_{\downarrow}$ is prohibited by time-reversal symmetry. Furthermore, mass terms $i\gamma_{\uparrow}\eta_{\downarrow}$ or $i\gamma_{\downarrow}\eta_{\uparrow}$ are forbidden because the system possesses an additional inversion symmetry $P_z$ (which is implemented non-trivially\cite{reuther14}). The gaplessness of all zero modes is, therefore, again symmetry-protected. In similarity to the $n_\uparrow=1$ case we again find local deviations from an exponential decrease in Fig.~\ref{fig:bound_modes}(c). Here, however, we could not identify a simple oscillating modulation that explains this behavior, possibly because interferences between the two Majorana modes at each vison complicate the situation compared to the $n_\uparrow=1$ phase.

The binding between flux excitations and fractional spin excitations has previously been described in exactly solvable Kitaev models on tri-coordinated lattices.\cite{knolle15,rachel16,theveniaut17} Furthermore, on the level of non-interacting fermion systems, Majorana zero modes bound to flux-vortex cores are a well known property of topological $p+ip$ superconductors.\cite{alicea12,kopnin91,read00,gurarie07,cheng09} Indeed, there is an exact mapping between the latter situation and the visons in our BZH-like spin liquid as we will demonstrate in the following. To show this equivalence, we go back to a real-space representation of Eq.~(\ref{eq:sl_bhz}) and assume that the system is in the phase with $n_\uparrow=1$. The general form of the Hamiltonian is then given by
\be
H=\sum_\pair\left[ \hat\Psi_\rr^\dagger
\sigma_\rrs^z\begin{pmatrix}
h_\rrs & 0\\ 0 & h^*_\rrs
\end{pmatrix}\hat\Psi_\rs+\text{H.c.}\right]\;.\label{bhz_real_sp}
\ee
Here $\hat{\Psi}_\rr=(f_{\uparrow\rr},f_{\uparrow\rr}^\dagger,f_{\downarrow\rr},f_{\downarrow\rr}^\dagger)^\text{T}$ and $\sigma^z_\rrs$ are the fluctuating gauge fields. We first consider the vison-free system (i.e. $\sigma^z_\rrs=1$ on all bonds) and derive a continuum version by expanding Eq.~(\ref{eq:sl_bhz2}) around $\kk=0$. This yields
\be
H=\int d^2\rr \hat\Psi^\dagger(\rr)
\begin{pmatrix}
h(\rr) & 0\\ 0 & h^*(\rr)
\end{pmatrix}\hat\Psi(\rr)\;,\label{bhz_real_sp_cont}
\ee
where
\be\label{eq:continuum}
h(\rr) = \begin{pmatrix}
-\frac{1}{2m}\nabla ^2 +\mu& \gamma \left( \partial_x + i\partial_y \right) \\
\gamma \left( \partial_x - i\partial_y \right) & \frac{1}{2m}\nabla ^2-\mu
\end{pmatrix}\;,
\ee
and $m=-\frac{1}{\alpha+\beta}$, $\mu=2\alpha+\beta$. Note that $\kk=0$ is a point of band inversion (negative mass in the upper band of $h(\rr)$ and positive mass in the lower band) such that Eq.~(\ref{eq:continuum}) correctly captures the topological properties of Eq.~(\ref{eq:sl_bhz2}) in the $n_\uparrow=1$ phase. This model describes a standard topological $p+ip$ superconductor with a uniform (and real) superconducting phase $\gamma$.
\begin{figure}
\includegraphics[width=0.7\linewidth]{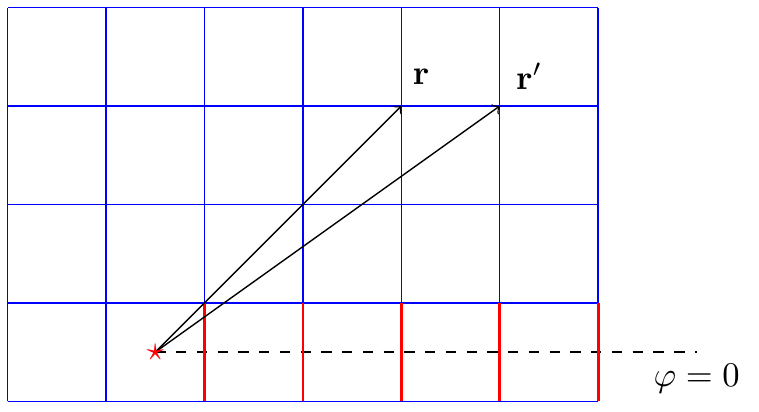}
\caption{Possible gauge-field configuration $\sigma_\rrs^z$ for a single vison, located at the origin (marked by a red star). Blue (red) bonds correspond to $\sigma_\rrs^z=1$ ($\sigma_\rrs^z=-1$). Bonds with $\sigma_\rrs^z=-1$ form a string along the $\varphi=0$ line. For large $\rr,\rs\gg1$, the angle $\varphi_\rr-\varphi_\rs$ between two coupled lattice sites $\rr$, $\rs$ vanishes, see text for details.}
\label{fig:vison_continuum}
\end{figure}

We now add a vison, located at the origin $\rr=0$ of the $x$-$y$-plane. In this case, the derivation of a continuum model requires some caution because the line of flipped gauge fields becomes a singular branch cut in the continuum limit. To define the exact gauge field configuration and to avoid such singularities we again go back to the discrete lattice version.  Assuming that the site positions are given by $\rr=(0.5+n_x, 0.5+n_y)$ with $n_x,n_y\in\mathds{Z}$ (such that the origin is located in the center of a unit square) a single vison at $\rr=0$ can be created by setting $\sigma^z_\rrs=-1$ on all bonds $(\rr,\rs)$ crossing the {\it positive} $x$-axis, while $\sigma^z_\rrs=1$ on all other bonds, see Fig.~\ref{fig:vison_continuum}. Note that the second vison is assumed to be infinitely far away. Next, we perform the gauge transformation
\be
\left(\begin{array}{c}
f_{\rr\uparrow}\\
f_{\rr\downarrow}^\dagger
\end{array}\right)\rightarrow 
\left(\begin{array}{cc}
e^{i\varphi_\rr/2} & 0 \\
0 & e^{-i\varphi_\rr/2}
\end{array}\right)
\left(\begin{array}{c}
f_{\rr\uparrow}\\
f_{\rr\downarrow}^\dagger
\end{array}\right)\;,\label{gauge_trafo_vison}
\ee 
where $\varphi_\rr\in[0,2\pi)$ is the polar angle of the vector $\rr$ in the $x$-$y$-plane (as usual, $\varphi=0$ corresponds to the positive $x$-axis). This transformation changes $h_\rrs$ in Eq.~(\ref{bhz_real_sp}) according to
\be
h_\rrs\rightarrow \left(\begin{array}{cc}
h_\rrs^{1,1}e^{-i(\varphi_\rr-\varphi_\rs)/2} & h_\rrs^{1,2}e^{-i(\varphi_\rr+\varphi_\rs)/2} \\[0.1cm]
h_\rrs^{2,1}e^{i(\varphi_\rr+\varphi_\rs)/2} & h_\rrs^{2,2}e^{i(\varphi_\rr-\varphi_\rs)/2}
\end{array}\right)\;.\label{vison_gauge_trafo}
\ee
Here, the superscript indices denote the matrix entries of $h_\rrs$. A continuum model can now be derived without any branch-cut singularities. For all bonds $(\rr,\rs)$ that do not cross the positive $x$-axis, the differences $\varphi_\rr-\varphi_\rs$ vanish in the large distance limit $\rr,\rs\gg1$ and one obtains
\be
h_\rrs \rightarrow \left(\begin{array}{cc}
h_\rrs^{1,1} & h_\rrs^{1,2}e^{-i\varphi_\rr} \\
h_\rrs^{2,1}e^{i\varphi_\rr} & h_\rrs^{2,2}
\end{array}\right)\;.
\ee
For all bonds $(\rr,\rs)$ that cross the positive $x$-axis (say $y>0$, $y'<0$) we can write $\varphi_\rr=\delta\varphi$, $\varphi_\rs=2\pi-\delta\varphi$ with $\delta\varphi>0$. In the continuum limit $\delta\varphi$ vanishes such that the effect of the gauge transformation on these bonds is given by
\be
h_\rrs\rightarrow-h_\rrs\;.
\ee
This shows that the gauge transformation exactly cancels the flipped gauge fields $\sigma^z_\rrs$ along the gauge string. Combining Eqs.~(\ref{eq:continuum}) and (\ref{vison_gauge_trafo}), a continuum model for a single vison at the origin can now be written as
\be\label{eq:continuum_vison}
h(\rr) = \begin{pmatrix}
-\frac{1}{2m}\nabla ^2 +\mu& \gamma \left( \partial_x + i\partial_y \right)e^{i\varphi_\rr} \\
\gamma \left( \partial_x - i\partial_y \right)e^{-i\varphi_\rr} & \frac{1}{2m}\nabla ^2-\mu
\end{pmatrix}\;.
\ee
Due to the phase factor $e^{i\varphi_\rr}$ winding around the origin, this is exactly the Bogoliubov-de Gennes Hamiltonian of a $p+ip$ superconductor with a single point-like flux vortex.\cite{cheng09,alicea12} It is well known that for this model topologically protected Majorana zero modes appear as gapless excitations in the vortex cores.

In the phase with $n_\uparrow=2$, an expansion of Eq.~(\ref{eq:sl_bhz}) around $\kk=0$ does not capture the full topology of the spinon bands, since $\kk=(\pi,\pi)$ is another point of inverted bands. In this case, an expansion of Eq.~(\ref{eq:sl_bhz}) around $\kk=(\pi,\pi)$ results, in total, in two copies of the model (\ref{eq:continuum_vison}), binding four Majorana zero modes in each vortex core. We therefore conclude that for a time-reversal invariant spin liquid with spinon-band Chern numbers $n_\uparrow$ and $n_\downarrow=-n_\uparrow$, there can be up to $|2n_\uparrow|$ Majorana modes (including both spin directions) tied to each vison excitation. Additional symmetries (such as $P_z$ in our case) can prevent them from gapping out each other. How the inclusion of dynamic gauge fields $\sim \sigma^x_\rrs$ modifies this observation remains a subject for future studies.
\begin{figure*}
\includegraphics[width=0.9\linewidth]{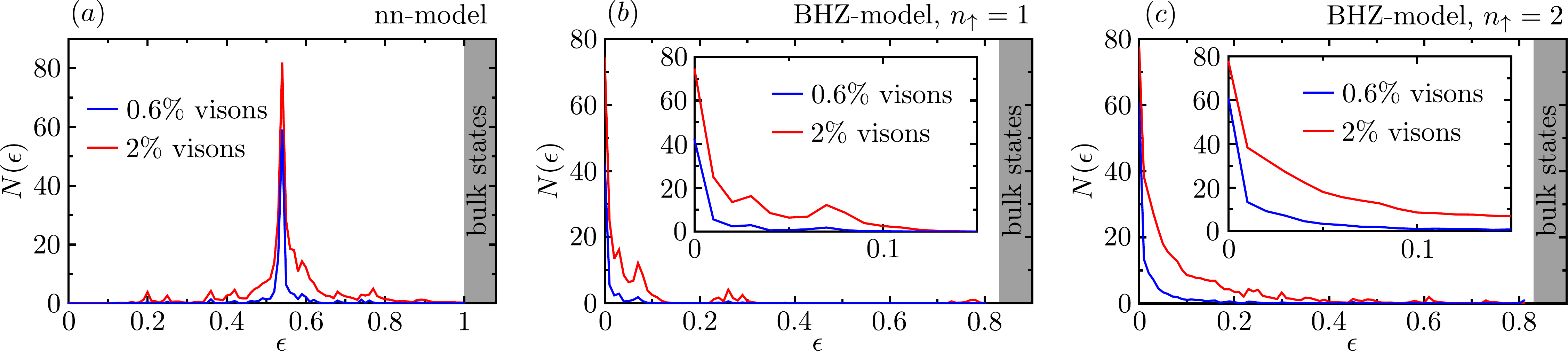}
\caption{Fermionic spinon density of states $N(\epsilon)$ in the bulk gap for a dilute gas of randomly arranged visons. The vison density is indicated for each curve. Using a system size of $100\times100$ lattice sites, $N(\epsilon)$ counts the number of states in each energy interval of length 0.01. All results are averaged over 100 different realizations. The thick gray line indicates the onset of the bulk gap. The data in (a) - (c) corresponds to the nearest-neighbor spin liquid, the BHZ-like model with $n_\uparrow=1$ and the BHZ-like model with $n_\uparrow=2$, respectively, with the same parameters as in Fig.~\ref{fig:vison_separation}. The insets in (b) and (c) show the density of states in the low-energy region.}
\label{fig:vison_gas}
\end{figure*}

\subsection{Vison gas}\label{vison_gas}
Having discussed the properties of a single vison pair, we finally consider the more realistic case where visons form a dilute gas of randomly arranged quasiparticles. Such a situation is, for instance, expected for thermally excited visons at finite temperatures or when visons bind to local lattice defects. We particularly investigate to which degree spinon-vison bound states at different vison cores hybridize and how such effects can be used for an experimental detection of topological spinon band-structures. To this end, we consider the three spin-liquid phases of the last sections and populate the lattice with a mean density of $0.6\%$ and $2\%$ randomly located visons, corresponding to an average vison distance of $\sim14$ and $\sim 7.5$ lattice spacings, respectively (note that a vison density of $x$ percent means that $x$ out of 100 elementary square plaquettes carry a vison). We then compute the spinon density of states inside the bulk gap and average the result over 100 different configurations. To avoid unwanted boundary effects, all calculations are performed on a torus.

For all three models we observe a narrow peak in the density of states at a position that coincides with the energies of the bound states for a single vison pair in Fig.~\ref{fig:bound_modes}. For the nearest-neighbor topologically trivial model [Fig.~\ref{fig:vison_gas}(a)] this peak is located at a finite energy inside the bulk gap, while for the BHZ-like model [Fig.~\ref{fig:vison_gas}(b), (c)] the density of states is maximal at zero energy. The fact that these maxima remain very narrow even for visons that are (on average) only a few lattice spacings apart, follows from the sharply peaked nature of the bound-state wave function, as shown in Fig.~\ref{fig:bound_modes}(d). Only for the BHZ-like model in the $n_\uparrow=2$ phase with $2\%$ visons (which in Fig.~\ref{fig:vison_gas} has the largest number of in-gap states), hybridization effects become more important and the bound modes start leaking into larger portions of the bulk gap. Generally, the density of states is roughly linear in the vison density and a Chern number $n_\uparrow=2$ additionally doubles the in-gap modes as compared to the $n_\uparrow=1$ phase. These observations might have interesting implications for experiments. We propose that topological spinon bands forming low-energy bound-state peaks in the spin-structure factor could be detectable in elastic neutron-scattering experiments. A plausible scenario would be that with increasing temperature (i.e., vison density) these peaks first become higher, since more bound states appear at zero energy. Above a certain temperature, the peaks would start to broaden since hybridization between the modes becomes stronger. Regardless of the spinon band-topology, the formation of bound states inside the spin gap would manifest in a shift of spectral weight from larger to smaller energies as the temperature increases.

Apart from the dominant peaks, Fig.~\ref{fig:vison_gas} also shows various smaller maxima such as the ones at $\epsilon\approx0.26$ in Fig.~\ref{fig:vison_gas}(b). These features appear if, by chance, two visons happen to be located very close to each other. Indeed, we find that the peaks at $\epsilon\approx0.26$ occur for a vison separation of roughly one lattice spacing. Taking into account dynamical gauge fields, we expect that these signatures might change significantly. For example, if vison hopping is allowed and the visons experience an attractive force at small distances [such as in Fig.~\ref{fig:vison_separation}(d)] the probability of finding two visons close together would increase, leading to higher secondary peaks. Otherwise, repulsive forces would diminish them.

\section{Discussion and conclusion}\label{conclusions}
In this work, we have investigated $\mathds{Z}_2$ spin liquids on the square lattice when SU(2) spin-rotation symmetry is maximally lifted. Spin liquids on their own are fascinating quantum states since they are examples for condensed matter realizations of gauge theories, with all their subtle implications for fractional quasiparticle excitations and topological ground-state degeneracies. Breaking spin-rotation symmetry adds another interesting aspect to these states as it allows one to construct spin phases where fractional spinon excitations exhibit topological band structures. The resulting spin liquids may be considered as the strongly coupled analogs of topological insulators or topological superconductors. A recent PSG classification of $\mathds{Z}_2$ spin liquids on the square lattice shows that topological band structures are indeed a generic property of spinons when SU(2) spin symmetry is lifted. Yet, the vast number of different spin phases and spinon band structures complicates their analysis enormously.

The motivation of this work is to reduce the complexity of the full set of PSG solutions by identifying the simplest possible anisotropic $\mathds{Z}_2$ spin liquids and investigate their spinon and vison excitations. Limiting the range of hopping and pairing amplitudes of the PSG mean-field ans\"atze, we find that for nearest-neighbor couplings only, a $\mathds{Z}_2$ gauge structure is incompatible with topological spinon bands. Extending the range of mean-field amplitudes to second neighbors and imposing certain simplifying assumptions on the structure of the PSG solutions (implementation of time reversal with $\mathcal{T}^2=-1$, block diagonal structure, and spin-isotropic second neighbor terms), only two spin-liquid solutions remain. Both have a form similar to the BHZ model for topological insulators. Particularly, the spinon bands are always topologically non-trivial, independent of the choice of the mean-field parameters.

We have selected three different mean-field models (one nearest-neighbor model and two BHZ-like models with Chern numbers $n_\uparrow=1$ and $n_\uparrow=2$ in one spin sector) and studied the properties of vison excitations. The problem simplifies considerably when we assume static gauge fields since for each fixed gauge-field configuration a free fermion system needs to be solved. We find that forces between pairs of visons are of very short-range type and die out after a few lattice spacings. Furthermore, for models with a topological spinon band structure we observe bound states between spinons and visons where $2|n_\uparrow|$ spinon-Majorana zero modes couple to each vison. The existence of these modes can be explained by mapping the spinon-vison system to a $p+ip$ superconductor with vortices. In the latter case, Majorana zero modes residing in the cores of superconducting vortices are a well established property.

In total, our analysis shows that the current level of approximation (i.e., including gauge fluctuations but neglecting their dynamics) still allows one to calculate vison-vison pair potentials as well as spinon-vison interaction effects. Concerning the latter, a vison excitation may be viewed as a point-like attracting potential for the spinons, trapping a small number of spinon modes. Other many-body effects can only be studied when dynamical gauge fields are considered. For example, dynamical gauge fluctuations could mediate short-range forces between the spinons, possibly leading to weakly coupled spinon-spinon bound states. This is in contrast to our mean-field treatment which assumes that spinons are, by construction, free fermionic objects that do not interact. A static approximation is also incompatible with the formation of vison-vison bound states for which the proper inclusion of vison kinetics is needed. Finally, it would be interesting to study the fate of topological spinon edge states when dynamical gauge fluctuations are considered. All these questions, however, require the solution of complicated many-body systems which is beyond the scope of the current work.

Concerning the experimental detection of spin liquids with topological spinon bands, our results indicate that spinon-vison Majorana modes form a narrow band at small energies even when the average vison distance is only a few lattice spacings. Such effects could be seen in the spin-structure factor measured in neutron scattering experiments. The small hybridization between different spinon-vison bound states stems from the strongly peaked nature of the corresponding wave functions. Whether this observation is more general and applies to larger classes of spin liquids remains a subject for future studies.

%====================================================================
\section{Acknowledgements}
We thank Jason Alicea, Johannes Knolle, Felix von Oppen, and Stephan Rachel for fruitful discussions. This work is supported by the DFG within the Transregio CRC 183 (project A02) and by the Freie Universit\"at Berlin within the Excellence Initiative of the German Research Foundation.
%====================================================================

\appendix
\section{Projective symmetry conditions on the mean-field amplitudes $U_\rrs$}\label{app_a}
In this appendix, we list the constraints on the mean-field parameters in all channels $u_\rrs^X$ with $X=s,t_1,t_2,t_3$ [see Eq.~(\ref{block})] following from a projective implementation of symmetries. In each channel $X$ the mean fields can be written as a function of $u_\dr^X$ which only depends on the {\it distance} $\drr\equiv(\dx,\dy)=\rs-\rr$ between sites $\rr$ and $\rs$,
\be
u_\rrs^X=\eta^{x\dy} u^X_\dr\;.
\ee
With this convention and Eq.~(\ref{eta}), the symmetry conditions on $u^X_\rrs$ read
\begin{eqnarray}
&-\eta_\T^{\dx+\dy}g_\T^\dagger u^s_\drr g_\T = u^s_\drr\;,&\notag\\
&\eta^\dx_1\eta^\dy_2 g_{P_x}^\dagger u^s_{P_x(\drr)} g_{P_x} = u^s_\drr\;,&\notag\\
&\eta^\dx_2\eta^\dy_1 g_{P_y}^\dagger u^s_{P_y(\drr)} g_{P_y} = u^s_\drr\;,&\notag\\
&\eta^{\dx \dy}g_{P_{xy}}^\dagger u^s_{P_{xy}(\drr)} g_{P_{xy}}=u^s_\drr\;,&\notag\\
&\eta_z^{\dx+\dy}g_{P_z}^\dagger u_\drr^s g_{P_z}=u_\drr^s\;,&\notag\\
&\eta^{\dx \dy} \left(u_{-\drr}^{s}\right)^\dagger=u^s_\drr\;,&\label{conditions_s}
\end{eqnarray}
\begin{eqnarray}
&-\eta_\T^{\dx+\dy}g_\T^\dagger u_\drr^{t_1} g_\T = u_\drr^{t_1}\;,&\notag\\
&-\eta^\dx_1\eta^\dy_2 g_{P_x}^\dagger u_{P_x(\drr)}^{t_1} g_{P_x} = u_\drr^{t_1}\;,&\notag\\
&-\eta^\dx_2\eta^\dy_1 g_{P_y}^\dagger u_{P_y(\drr)}^{t_1} g_{P_y} = u_\drr^{t_1}\;,&\notag\\
&-\eta^{\dx\dy}g_{P_{xy}}^\dagger u_{P_{xy}(\drr)}^{t_1} g_{P_{xy}}=u_\drr^{t_1}\;,&\notag\\
&\eta_z^{\dx+\dy}g_{P_z}^\dagger u_\drr^{t_1} g_{P_z}=u_\drr^{t_1}\;,&\notag\\
&\eta^{\dx\dy} \left(u_{-\drr}^{t_1}\right)^\dagger =u_\drr^{t_1}\;,&\label{conditions_t1}
\end{eqnarray}
\begin{eqnarray}
&-\eta_\T^{\dx+\dy}g_\T^\dagger u_\drr^{t_2} g_\T = u_\drr^{t_2}\;,&\notag\\
&-\eta^{\dx}_1\eta^{\dy}_2 g_{P_x}^\dagger u_{P_x(\drr)}^{t_2} g_{P_x} = u_\drr^{t_2}\;,&\notag\\
&\eta^{\dx}_2\eta^{\dy}_1 g_{P_y}^\dagger u_{P_y(\drr)}^{t_2} g_{P_y} = u_\drr^{t_2}\;,&\notag\\
&-i\eta^{\dx \dy}g_{P_{xy}}^\dagger u_{P_{xy}(\drr)}^{t_2} g_{P_{xy}}=u_\drr^{t_3}\;,&\notag\\
&-\eta_z^{\dx+\dy}g_{P_z}^\dagger u_\drr^{t_2} g_{P_z}=u_\drr^{t_2}\;,&\notag\\
&-\eta^{\dx \dy} \left(u_{-\drr}^{t_2}\right)^\dagger =u_\drr^{t_2}\;,&\label{conditions_t2}
\end{eqnarray}
and
\begin{eqnarray}
&-\eta_\T^{\dx+\dy}g_\T^\dagger u_\drr^{t_3} g_\T = u_\drr^{t_3}\;,&\notag\\
&\eta^{\dx}_1\eta^{\dy}_2 g_{P_x}^\dagger u_{P_x(\drr)}^{t_3} g_{P_x} = u_\drr^{t_3}\;,&\notag\\
&-\eta^{\dx}_2\eta^{\dy}_1 g_{P_y}^\dagger u_{P_y(\drr)}^{t_3} g_{P_y} = u_\drr^{t_3}\;,&\notag\\
&i\eta^{\dx \dy}g_{P_{xy}}^\dagger u_{P_{xy}(\drr)}^{t_3} g_{P_{xy}}=u_\drr^{t_2}\;,&\notag\\
&-\eta_z^{\dx+\dy}g_{P_z}^\dagger u_\drr^{t_3} g_{P_z}=u_\drr^{t_3}\;,&\notag\\
&\eta^{\dx \dy} \left(u_{-\drr}^{t_3}\right)^\dagger =u_\drr^{t_3}\;.&\label{conditions_t3}
\end{eqnarray}
In each of these equations, the last line ensures hermiticity of the mean-field Hamiltonian.

\section{$\mathds{Z}_2$ PSG representations with finite nearest neighbor mean-field parameters}\label{app_b}
Restricting the range of PSG mean-field amplitudes to nearest neighbors leads to 28 finite $\mathds{Z}_2$ solutions with broken SU(2) spin-rotation symmetry. Labeling these solutions by the matrices $g_{P_z}$, $g_{\mathcal{T}}$, $g_{P_{xy}}$, $g_{P_x}$, $g_{P_y}$ and the sign parameters $\eta_z$, $\eta_\mathcal{T}$, $\eta$, $\eta_1$, $\eta_2$ [see Eq.~(\ref{eta})] one finds that they come in pairs which only differ by the parameter $\eta$. Here, we provide the full list of such PSG solutions (grouped in pairs with $\eta=\pm 1$),
\begin{eqnarray}
&g_{P_z}=\sigma^0, \quad g_{\mathcal{T}}=\sigma^0, \quad g_{P_{xy}}=i\sigma^3,&\notag\\
&g_{P_x}=i\sigma^1, \quad g_{P_y}=i\sigma^1,&\notag\\
&\eta_z=-1, \quad \eta_{\mathcal{T}}=-1, \quad \eta=\pm 1,\quad\eta_1=1, \quad \eta_2=1&\notag\\
\end{eqnarray}
\begin{eqnarray}
&g_{P_z}=\sigma^0, \quad g_{\mathcal{T}}=\sigma^0, \quad g_{P_{xy}}=i\sigma^3, &\notag\\
&g_{P_x}=i\sigma^1, \quad g_{P_y}=i\sigma^1,&\notag\\
&\eta_z=-1, \quad \eta_{\mathcal{T}}=-1, \quad \eta=\pm 1, \quad\eta_1=-1, \quad \eta_2=-1&\notag\\
\end{eqnarray}
\begin{eqnarray}
&g_{P_z}=i\sigma^3, \quad g_{\mathcal{T}}=\sigma^0, \quad g_{P_{xy}}=i\sigma^3, &\notag\\
&g_{P_x}=i\sigma^1, \quad g_{P_y}=i\sigma^1,&\notag\\
&\eta_z=1, \quad \eta_{\mathcal{T}}=-1, \quad \eta=\pm 1,\quad\eta_1=-1, \quad \eta_2=-1&\notag\\
\end{eqnarray}
\begin{eqnarray}
&g_{P_z}=i\sigma^2, \quad g_{\mathcal{T}}=\sigma^0, \quad g_{P_{xy}}=\sigma^0, &\notag\\
&g_{P_x}=i\sigma^3, \quad g_{P_y}=i\sigma^3,&\notag\\
&\eta_z=1, \quad \eta_{\mathcal{T}}=-1, \quad \eta=\pm 1,\quad\eta_1=-1, \quad \eta_2=-1&\notag\\
\end{eqnarray}
\begin{eqnarray}
&g_{P_z}=i\sigma^2, \quad g_{\mathcal{T}}=\sigma^0, \quad g_{P_{xy}}=i\sigma^3, &\notag\\
&g_{P_x}=\sigma^0, \quad g_{P_y}=\sigma^0,&\notag\\
&\eta_z=1, \quad \eta_{\mathcal{T}}=-1, \quad \eta=\pm 1,\quad\eta_1=1, \quad \eta_2=1&\notag\\
\end{eqnarray}
\begin{eqnarray}
&g_{P_z}=i\sigma^1, \quad g_{\mathcal{T}}=\sigma^0, \quad g_{P_{xy}}=i\sigma^3, &\notag\\
&g_{P_x}=i\sigma^1, \quad g_{P_y}=i\sigma^1,&\notag\\
&\eta_z=1, \quad \eta_{\mathcal{T}}=-1, \quad \eta=\pm 1,\quad\eta_1=1, \quad \eta_2=1&\notag\\
\end{eqnarray}
\begin{eqnarray}
&g_{P_z}=i\sigma^2, \quad g_{\mathcal{T}}=\sigma^0, \quad g_{P_{xy}}=i\sigma^3, &\notag\\
&g_{P_x}=\sigma^0, \quad g_{P_y}=\sigma^0,&\notag\\
&\eta_z=-1, \quad \eta_{\mathcal{T}}=-1, \quad \eta=\pm 1,\quad\eta_1=1, \quad \eta_2=-1&\notag\\
\end{eqnarray}
\begin{eqnarray}
&g_{P_z}=i\sigma^1, \quad g_{\mathcal{T}}=\sigma^0, \quad g_{P_{xy}}=i\sigma^3, &\notag\\
&g_{P_x}=i\sigma^1, \quad g_{P_y}=i\sigma^1,&\notag\\
&\eta_z=-1, \quad \eta_{\mathcal{T}}=-1, \quad \eta=\pm 1,\quad\eta_1=-1, \quad \eta_2=1&\notag\\
\end{eqnarray}
\begin{eqnarray}
&g_{P_z}=i\sigma^2, \quad g_{\mathcal{T}}=\sigma^0, \quad g_{P_{xy}}=i\sigma^3, &\notag\\
&g_{P_x}=i\sigma^3, \quad g_{P_y}=i\sigma^3,&\notag\\
&\eta_z=1, \quad \eta_{\mathcal{T}}=-1, \quad \eta=\pm 1,\quad\eta_1=-1, \quad \eta_2=-1&\notag\\
\end{eqnarray}
\begin{eqnarray}
&g_{P_z}=i\sigma^2, \quad g_{\mathcal{T}}=\sigma^0, \quad g_{P_{xy}}=i\sigma^3, &\notag\\
&g_{P_x}=i\sigma^1, \quad g_{P_y}=i\sigma^1,&\notag\\
&\eta_z=1, \quad \eta_{\mathcal{T}}=-1, \quad \eta=\pm 1,\quad\eta_1=-1, \quad \eta_2=-1&\notag\\
\end{eqnarray}
\begin{eqnarray}
&g_{P_z}=i\sigma^1, \quad g_{\mathcal{T}}=\sigma^0, \quad g_{P_{xy}}=i\sigma^3, &\notag\\
&g_{P_x}=i\sigma^1, \quad g_{P_y}=i\sigma^1,&\notag\\
&\eta_z=-1, \quad \eta_{\mathcal{T}}=-1, \quad \eta=\pm 1,\quad\eta_1=-1, \quad \eta_2=-1&\notag\\
\end{eqnarray}
\begin{eqnarray}
&g_{P_z}=i\sigma^2, \quad g_{\mathcal{T}}=\sigma^0, \quad g_{P_{xy}}=i\sigma^3, &\notag\\
&g_{P_x}=\sigma^0, \quad g_{P_y}=\sigma^0,&\notag\\
&\eta_z=-1, \quad \eta_{\mathcal{T}}=-1, \quad \eta=\pm 1,\quad\eta_1=1, \quad \eta_2=1&\notag\\
\end{eqnarray}
\begin{eqnarray}
&g_{P_z}=i\sigma^2, \quad g_{\mathcal{T}}=\sigma^0, \quad g_{P_{xy}}=i\sigma^3, &\notag\\
&g_{P_x}=\sigma^0, \quad g_{P_y}=\sigma^0,&\notag\\
&\eta_z=1, \quad \eta_{\mathcal{T}}=-1, \quad \eta=\pm 1,\quad\eta_1=-1, \quad \eta_2=-1&\notag\\
\end{eqnarray}
\begin{eqnarray}
&g_{P_z}=i\sigma^1, \quad g_{\mathcal{T}}=\sigma^0, \quad g_{P_{xy}}=i\sigma^3, &\notag\\
&g_{P_x}=i\sigma^1, \quad g_{P_y}=i\sigma^1,&\notag\\
&\eta_z=1, \quad \eta_{\mathcal{T}}=-1, \quad \eta=\pm 1, 
\quad\eta_1=-1, \quad \eta_2=-1&\notag\\
\end{eqnarray}


\begin{thebibliography}{75}%
\makeatletter
\providecommand \@ifxundefined [1]{%
 \@ifx{#1\undefined}
}%
\providecommand \@ifnum [1]{%
 \ifnum #1\expandafter \@firstoftwo
 \else \expandafter \@secondoftwo
 \fi
}%
\providecommand \@ifx [1]{%
 \ifx #1\expandafter \@firstoftwo
 \else \expandafter \@secondoftwo
 \fi
}%
\providecommand \natexlab [1]{#1}%
\providecommand \enquote  [1]{``#1''}%
\providecommand \bibnamefont  [1]{#1}%
\providecommand \bibfnamefont [1]{#1}%
\providecommand \citenamefont [1]{#1}%
\providecommand \href@noop [0]{\@secondoftwo}%
\providecommand \href [0]{\begingroup \@sanitize@url \@href}%
\providecommand \@href[1]{\@@startlink{#1}\@@href}%
\providecommand \@@href[1]{\endgroup#1\@@endlink}%
\providecommand \@sanitize@url [0]{\catcode `\\12\catcode `\$12\catcode
  `\&12\catcode `\#12\catcode `\^12\catcode `\_12\catcode `\%12\relax}%
\providecommand \@@startlink[1]{}%
\providecommand \@@endlink[0]{}%
\providecommand \url  [0]{\begingroup\@sanitize@url \@url }%
\providecommand \@url [1]{\endgroup\@href {#1}{\urlprefix }}%
\providecommand \urlprefix  [0]{URL }%
\providecommand \Eprint [0]{\href }%
\providecommand \doibase [0]{http://dx.doi.org/}%
\providecommand \selectlanguage [0]{\@gobble}%
\providecommand \bibinfo  [0]{\@secondoftwo}%
\providecommand \bibfield  [0]{\@secondoftwo}%
\providecommand \translation [1]{[#1]}%
\providecommand \BibitemOpen [0]{}%
\providecommand \bibitemStop [0]{}%
\providecommand \bibitemNoStop [0]{.\EOS\space}%
\providecommand \EOS [0]{\spacefactor3000\relax}%
\providecommand \BibitemShut  [1]{\csname bibitem#1\endcsname}%
\let\auto@bib@innerbib\@empty
%</preamble>
\bibitem [{\citenamefont {Balents}(2010)}]{balents10}%
  \BibitemOpen
  \bibfield  {author} {\bibinfo {author} {\bibfnamefont {L.}~\bibnamefont
  {Balents}},\ }\href {http://dx.doi.org/10.1038/nature08917} {\bibfield
  {journal} {\bibinfo  {journal} {Nature}\ }\textbf {\bibinfo {volume} {464}},\
  \bibinfo {pages} {199} (\bibinfo {year} {2010})}\BibitemShut {NoStop}%
\bibitem [{\citenamefont {Savary}\ and\ \citenamefont
  {Balents}(2016)}]{savary16}%
  \BibitemOpen
  \bibfield  {author} {\bibinfo {author} {\bibfnamefont {L.}~\bibnamefont
  {Savary}}\ and\ \bibinfo {author} {\bibfnamefont {L.}~\bibnamefont
  {Balents}},\ }\href {http://stacks.iop.org/0034-4885/80/i=1/a=016502}
  {\bibfield  {journal} {\bibinfo  {journal} {Reports on Progress in Physics}\
  }\textbf {\bibinfo {volume} {80}},\ \bibinfo {pages} {016502} (\bibinfo
  {year} {2016})}\BibitemShut {NoStop}%
\bibitem [{\citenamefont {Lee}(2008)}]{lee08}%
  \BibitemOpen
  \bibfield  {author} {\bibinfo {author} {\bibfnamefont {P.~A.}\ \bibnamefont
  {Lee}},\ }\href {\doibase 10.1126/science.1163196} {\bibfield  {journal}
  {\bibinfo  {journal} {Science}\ }\textbf {\bibinfo {volume} {321}},\ \bibinfo
  {pages} {1306} (\bibinfo {year} {2008})}\BibitemShut {NoStop}%
\bibitem [{\citenamefont {Anderson}(1973)}]{anderson73}%
  \BibitemOpen
  \bibfield  {author} {\bibinfo {author} {\bibfnamefont {P.~W.}\ \bibnamefont
  {Anderson}},\ }\href@noop {} {\bibfield  {journal} {\bibinfo  {journal}
  {Materials Research Bulletin}\ }\textbf {\bibinfo {volume} {8}},\ \bibinfo
  {pages} {153} (\bibinfo {year} {1973})}\BibitemShut {NoStop}%
\bibitem [{\citenamefont {Wen}(2007)}]{wen_book}%
  \BibitemOpen
  \bibfield  {author} {\bibinfo {author} {\bibfnamefont {X.-G.}\ \bibnamefont
  {Wen}},\ }\href@noop {} {\emph {\bibinfo {title} {Quantum Field Theory of
  Many-Body Systems: From the Origin of Sound to an Origin of Light and
  Electrons}}}\ (\bibinfo  {publisher} {Oxford Graduate Texts},\ \bibinfo
  {year} {2007})\BibitemShut {NoStop}%
\bibitem [{\citenamefont {Lacroix}\ \emph {et~al.}(2011)\citenamefont
  {Lacroix}, \citenamefont {Mendels},\ and\ \citenamefont
  {Mila}}]{frustration_book}%
  \BibitemOpen
  \bibfield  {author} {\bibinfo {author} {\bibfnamefont {C.}~\bibnamefont
  {Lacroix}}, \bibinfo {author} {\bibfnamefont {P.}~\bibnamefont {Mendels}}, \
  and\ \bibinfo {author} {\bibfnamefont {F.}~\bibnamefont {Mila}},\ }\href@noop
  {} {\emph {\bibinfo {title} {Introduction to Frustrated Magnetism}}}\
  (\bibinfo  {publisher} {Springer},\ \bibinfo {year} {2011})\BibitemShut
  {NoStop}%
\bibitem [{\citenamefont {Wegner}(1971)}]{wegner71}%
  \BibitemOpen
  \bibfield  {author} {\bibinfo {author} {\bibfnamefont {F.~J.}\ \bibnamefont
  {Wegner}},\ }\href {\doibase 10.1063/1.1665530} {\bibfield  {journal}
  {\bibinfo  {journal} {J. Math. Phys.}\ }\textbf {\bibinfo {volume} {12}},\
  \bibinfo {pages} {2259} (\bibinfo {year} {1971})}\BibitemShut {NoStop}%
\bibitem [{\citenamefont {Kogut}(1979)}]{kogut79}%
  \BibitemOpen
  \bibfield  {author} {\bibinfo {author} {\bibfnamefont {J.~B.}\ \bibnamefont
  {Kogut}},\ }\href {\doibase 10.1103/RevModPhys.51.659} {\bibfield  {journal}
  {\bibinfo  {journal} {Rev. Mod. Phys.}\ }\textbf {\bibinfo {volume} {51}},\
  \bibinfo {pages} {659} (\bibinfo {year} {1979})}\BibitemShut {NoStop}%
\bibitem [{\citenamefont {Senthil}\ and\ \citenamefont
  {Fisher}(2000)}]{senthil00}%
  \BibitemOpen
  \bibfield  {author} {\bibinfo {author} {\bibfnamefont {T.}~\bibnamefont
  {Senthil}}\ and\ \bibinfo {author} {\bibfnamefont {M.~P.~A.}\ \bibnamefont
  {Fisher}},\ }\href {\doibase 10.1103/PhysRevB.62.7850} {\bibfield  {journal}
  {\bibinfo  {journal} {Phys. Rev. B}\ }\textbf {\bibinfo {volume} {62}},\
  \bibinfo {pages} {7850} (\bibinfo {year} {2000})}\BibitemShut {NoStop}%
\bibitem [{\citenamefont {Read}\ and\ \citenamefont {Sachdev}(1991)}]{read91}%
  \BibitemOpen
  \bibfield  {author} {\bibinfo {author} {\bibfnamefont {N.}~\bibnamefont
  {Read}}\ and\ \bibinfo {author} {\bibfnamefont {S.}~\bibnamefont {Sachdev}},\
  }\href {\doibase 10.1103/PhysRevLett.66.1773} {\bibfield  {journal} {\bibinfo
   {journal} {Phys. Rev. Lett.}\ }\textbf {\bibinfo {volume} {66}},\ \bibinfo
  {pages} {1773} (\bibinfo {year} {1991})}\BibitemShut {NoStop}%
\bibitem [{\citenamefont {Wen}(1991)}]{wen91}%
  \BibitemOpen
  \bibfield  {author} {\bibinfo {author} {\bibfnamefont {X.~G.}\ \bibnamefont
  {Wen}},\ }\href {\doibase 10.1103/PhysRevB.44.2664} {\bibfield  {journal}
  {\bibinfo  {journal} {Phys. Rev. B}\ }\textbf {\bibinfo {volume} {44}},\
  \bibinfo {pages} {2664} (\bibinfo {year} {1991})}\BibitemShut {NoStop}%
\bibitem [{\citenamefont {Read}\ and\ \citenamefont
  {Chakraborty}(1989)}]{read89}%
  \BibitemOpen
  \bibfield  {author} {\bibinfo {author} {\bibfnamefont {N.}~\bibnamefont
  {Read}}\ and\ \bibinfo {author} {\bibfnamefont {B.}~\bibnamefont
  {Chakraborty}},\ }\href {\doibase 10.1103/PhysRevB.40.7133} {\bibfield
  {journal} {\bibinfo  {journal} {Phys. Rev. B}\ }\textbf {\bibinfo {volume}
  {40}},\ \bibinfo {pages} {7133} (\bibinfo {year} {1989})}\BibitemShut
  {NoStop}%
\bibitem [{\citenamefont {Kivelson}(1989)}]{kivelson89}%
  \BibitemOpen
  \bibfield  {author} {\bibinfo {author} {\bibfnamefont {S.}~\bibnamefont
  {Kivelson}},\ }\href {\doibase 10.1103/PhysRevB.39.259} {\bibfield  {journal}
  {\bibinfo  {journal} {Phys. Rev. B}\ }\textbf {\bibinfo {volume} {39}},\
  \bibinfo {pages} {259} (\bibinfo {year} {1989})}\BibitemShut {NoStop}%
\bibitem [{\citenamefont {Punk}\ \emph {et~al.}(2014)\citenamefont {Punk},
  \citenamefont {Chowdhury},\ and\ \citenamefont {Sachdev}}]{punk14}%
  \BibitemOpen
  \bibfield  {author} {\bibinfo {author} {\bibfnamefont {M.}~\bibnamefont
  {Punk}}, \bibinfo {author} {\bibfnamefont {D.}~\bibnamefont {Chowdhury}}, \
  and\ \bibinfo {author} {\bibfnamefont {S.}~\bibnamefont {Sachdev}},\ }\href
  {http://dx.doi.org/10.1038/nphys2887} {\bibfield  {journal} {\bibinfo
  {journal} {Nat. Phys.}\ }\textbf {\bibinfo {volume} {10}},\ \bibinfo {pages}
  {289} (\bibinfo {year} {2014})}\BibitemShut {NoStop}%
\bibitem [{\citenamefont {Polyakov}(1977)}]{polyakov77}%
  \BibitemOpen
  \bibfield  {author} {\bibinfo {author} {\bibfnamefont {A.}~\bibnamefont
  {Polyakov}},\ }\href {\doibase
  http://dx.doi.org/10.1016/0550-3213(77)90086-4} {\bibfield  {journal}
  {\bibinfo  {journal} {Nuclear Physics B}\ }\textbf {\bibinfo {volume}
  {120}},\ \bibinfo {pages} {429 } (\bibinfo {year} {1977})}\BibitemShut
  {NoStop}%
\bibitem [{\citenamefont {Hermele}\ \emph {et~al.}(2004)\citenamefont
  {Hermele}, \citenamefont {Senthil}, \citenamefont {Fisher}, \citenamefont
  {Lee}, \citenamefont {Nagaosa},\ and\ \citenamefont {Wen}}]{hermele04}%
  \BibitemOpen
  \bibfield  {author} {\bibinfo {author} {\bibfnamefont {M.}~\bibnamefont
  {Hermele}}, \bibinfo {author} {\bibfnamefont {T.}~\bibnamefont {Senthil}},
  \bibinfo {author} {\bibfnamefont {M.~P.~A.}\ \bibnamefont {Fisher}}, \bibinfo
  {author} {\bibfnamefont {P.~A.}\ \bibnamefont {Lee}}, \bibinfo {author}
  {\bibfnamefont {N.}~\bibnamefont {Nagaosa}}, \ and\ \bibinfo {author}
  {\bibfnamefont {X.-G.}\ \bibnamefont {Wen}},\ }\href {\doibase
  10.1103/PhysRevB.70.214437} {\bibfield  {journal} {\bibinfo  {journal} {Phys.
  Rev. B}\ }\textbf {\bibinfo {volume} {70}},\ \bibinfo {pages} {214437}
  (\bibinfo {year} {2004})}\BibitemShut {NoStop}%
\bibitem [{\citenamefont {Ran}\ \emph {et~al.}(2009)\citenamefont {Ran},
  \citenamefont {Ko}, \citenamefont {Lee},\ and\ \citenamefont {Wen}}]{ran09}%
  \BibitemOpen
  \bibfield  {author} {\bibinfo {author} {\bibfnamefont {Y.}~\bibnamefont
  {Ran}}, \bibinfo {author} {\bibfnamefont {W.-H.}\ \bibnamefont {Ko}},
  \bibinfo {author} {\bibfnamefont {P.~A.}\ \bibnamefont {Lee}}, \ and\
  \bibinfo {author} {\bibfnamefont {X.-G.}\ \bibnamefont {Wen}},\ }\href
  {\doibase 10.1103/PhysRevLett.102.047205} {\bibfield  {journal} {\bibinfo
  {journal} {Phys. Rev. Lett.}\ }\textbf {\bibinfo {volume} {102}},\ \bibinfo
  {pages} {047205} (\bibinfo {year} {2009})}\BibitemShut {NoStop}%
\bibitem [{\citenamefont {Pesin}\ and\ \citenamefont
  {Balents}(2010)}]{pesin10}%
  \BibitemOpen
  \bibfield  {author} {\bibinfo {author} {\bibfnamefont {D.}~\bibnamefont
  {Pesin}}\ and\ \bibinfo {author} {\bibfnamefont {L.}~\bibnamefont
  {Balents}},\ }\href
  {http://www.nature.com/nphys/journal/v6/n5/abs/nphys1606.html} {\bibfield
  {journal} {\bibinfo  {journal} {Nat. Phys.}\ }\textbf {\bibinfo {volume}
  {6}},\ \bibinfo {pages} {376} (\bibinfo {year} {2010})}\BibitemShut {NoStop}%
\bibitem [{\citenamefont {R\"uegg}\ and\ \citenamefont
  {Fiete}(2012)}]{ruegg12}%
  \BibitemOpen
  \bibfield  {author} {\bibinfo {author} {\bibfnamefont {A.}~\bibnamefont
  {R\"uegg}}\ and\ \bibinfo {author} {\bibfnamefont {G.~A.}\ \bibnamefont
  {Fiete}},\ }\href {\doibase 10.1103/PhysRevLett.108.046401} {\bibfield
  {journal} {\bibinfo  {journal} {Phys. Rev. Lett.}\ }\textbf {\bibinfo
  {volume} {108}},\ \bibinfo {pages} {046401} (\bibinfo {year}
  {2012})}\BibitemShut {NoStop}%
\bibitem [{\citenamefont {Rachel}\ and\ \citenamefont
  {Le~Hur}(2010)}]{rachel10}%
  \BibitemOpen
  \bibfield  {author} {\bibinfo {author} {\bibfnamefont {S.}~\bibnamefont
  {Rachel}}\ and\ \bibinfo {author} {\bibfnamefont {K.}~\bibnamefont
  {Le~Hur}},\ }\href {\doibase 10.1103/PhysRevB.82.075106} {\bibfield
  {journal} {\bibinfo  {journal} {Phys. Rev. B}\ }\textbf {\bibinfo {volume}
  {82}},\ \bibinfo {pages} {075106} (\bibinfo {year} {2010})}\BibitemShut
  {NoStop}%
\bibitem [{\citenamefont {Cho}\ \emph {et~al.}(2012)\citenamefont {Cho},
  \citenamefont {Lu},\ and\ \citenamefont {Moore}}]{cho12}%
  \BibitemOpen
  \bibfield  {author} {\bibinfo {author} {\bibfnamefont {G.~Y.}\ \bibnamefont
  {Cho}}, \bibinfo {author} {\bibfnamefont {Y.-M.}\ \bibnamefont {Lu}}, \ and\
  \bibinfo {author} {\bibfnamefont {J.~E.}\ \bibnamefont {Moore}},\ }\href
  {\doibase 10.1103/PhysRevB.86.125101} {\bibfield  {journal} {\bibinfo
  {journal} {Phys. Rev. B}\ }\textbf {\bibinfo {volume} {86}},\ \bibinfo
  {pages} {125101} (\bibinfo {year} {2012})}\BibitemShut {NoStop}%
\bibitem [{\citenamefont {Kitaev}(2006)}]{kitaev06}%
  \BibitemOpen
  \bibfield  {author} {\bibinfo {author} {\bibfnamefont {A.}~\bibnamefont
  {Kitaev}},\ }\href {\doibase http://dx.doi.org/10.1016/j.aop.2005.10.005}
  {\bibfield  {journal} {\bibinfo  {journal} {Annals of Physics}\ }\textbf
  {\bibinfo {volume} {321}},\ \bibinfo {pages} {2 } (\bibinfo {year}
  {2006})}\BibitemShut {NoStop}%
\bibitem [{\citenamefont {Baskaran}\ \emph {et~al.}(2007)\citenamefont
  {Baskaran}, \citenamefont {Mandal},\ and\ \citenamefont
  {Shankar}}]{baskaran07}%
  \BibitemOpen
  \bibfield  {author} {\bibinfo {author} {\bibfnamefont {G.}~\bibnamefont
  {Baskaran}}, \bibinfo {author} {\bibfnamefont {S.}~\bibnamefont {Mandal}}, \
  and\ \bibinfo {author} {\bibfnamefont {R.}~\bibnamefont {Shankar}},\ }\href
  {\doibase 10.1103/PhysRevLett.98.247201} {\bibfield  {journal} {\bibinfo
  {journal} {Phys. Rev. Lett.}\ }\textbf {\bibinfo {volume} {98}},\ \bibinfo
  {pages} {247201} (\bibinfo {year} {2007})}\BibitemShut {NoStop}%
\bibitem [{\citenamefont {Knolle}\ \emph {et~al.}(2014)\citenamefont {Knolle},
  \citenamefont {Kovrizhin}, \citenamefont {Chalker},\ and\ \citenamefont
  {Moessner}}]{knolle14}%
  \BibitemOpen
  \bibfield  {author} {\bibinfo {author} {\bibfnamefont {J.}~\bibnamefont
  {Knolle}}, \bibinfo {author} {\bibfnamefont {D.~L.}\ \bibnamefont
  {Kovrizhin}}, \bibinfo {author} {\bibfnamefont {J.~T.}\ \bibnamefont
  {Chalker}}, \ and\ \bibinfo {author} {\bibfnamefont {R.}~\bibnamefont
  {Moessner}},\ }\href {\doibase 10.1103/PhysRevLett.112.207203} {\bibfield
  {journal} {\bibinfo  {journal} {Phys. Rev. Lett.}\ }\textbf {\bibinfo
  {volume} {112}},\ \bibinfo {pages} {207203} (\bibinfo {year}
  {2014})}\BibitemShut {NoStop}%
\bibitem [{\citenamefont {O'Brien}\ \emph {et~al.}(2016)\citenamefont
  {O'Brien}, \citenamefont {Hermanns},\ and\ \citenamefont {Trebst}}]{brien16}%
  \BibitemOpen
  \bibfield  {author} {\bibinfo {author} {\bibfnamefont {K.}~\bibnamefont
  {O'Brien}}, \bibinfo {author} {\bibfnamefont {M.}~\bibnamefont {Hermanns}}, \
  and\ \bibinfo {author} {\bibfnamefont {S.}~\bibnamefont {Trebst}},\ }\href
  {\doibase 10.1103/PhysRevB.93.085101} {\bibfield  {journal} {\bibinfo
  {journal} {Phys. Rev. B}\ }\textbf {\bibinfo {volume} {93}},\ \bibinfo
  {pages} {085101} (\bibinfo {year} {2016})}\BibitemShut {NoStop}%
\bibitem [{\citenamefont {Kells}\ and\ \citenamefont {Vala}(2010)}]{kells10}%
  \BibitemOpen
  \bibfield  {author} {\bibinfo {author} {\bibfnamefont {G.}~\bibnamefont
  {Kells}}\ and\ \bibinfo {author} {\bibfnamefont {J.}~\bibnamefont {Vala}},\
  }\href {\doibase 10.1103/PhysRevB.82.125122} {\bibfield  {journal} {\bibinfo
  {journal} {Phys. Rev. B}\ }\textbf {\bibinfo {volume} {82}},\ \bibinfo
  {pages} {125122} (\bibinfo {year} {2010})}\BibitemShut {NoStop}%
\bibitem [{\citenamefont {Thakurathi}\ \emph {et~al.}(2014)\citenamefont
  {Thakurathi}, \citenamefont {Sengupta},\ and\ \citenamefont
  {Sen}}]{thakurathi14}%
  \BibitemOpen
  \bibfield  {author} {\bibinfo {author} {\bibfnamefont {M.}~\bibnamefont
  {Thakurathi}}, \bibinfo {author} {\bibfnamefont {K.}~\bibnamefont
  {Sengupta}}, \ and\ \bibinfo {author} {\bibfnamefont {D.}~\bibnamefont
  {Sen}},\ }\href {\doibase 10.1103/PhysRevB.89.235434} {\bibfield  {journal}
  {\bibinfo  {journal} {Phys. Rev. B}\ }\textbf {\bibinfo {volume} {89}},\
  \bibinfo {pages} {235434} (\bibinfo {year} {2014})}\BibitemShut {NoStop}%
\bibitem [{\citenamefont {Hermanns}\ \emph {et~al.}(2015)\citenamefont
  {Hermanns}, \citenamefont {O'Brien},\ and\ \citenamefont
  {Trebst}}]{hermanns15}%
  \BibitemOpen
  \bibfield  {author} {\bibinfo {author} {\bibfnamefont {M.}~\bibnamefont
  {Hermanns}}, \bibinfo {author} {\bibfnamefont {K.}~\bibnamefont {O'Brien}}, \
  and\ \bibinfo {author} {\bibfnamefont {S.}~\bibnamefont {Trebst}},\ }\href
  {\doibase 10.1103/PhysRevLett.114.157202} {\bibfield  {journal} {\bibinfo
  {journal} {Phys. Rev. Lett.}\ }\textbf {\bibinfo {volume} {114}},\ \bibinfo
  {pages} {157202} (\bibinfo {year} {2015})}\BibitemShut {NoStop}%
\bibitem [{\citenamefont {Knolle}\ \emph {et~al.}(2015)\citenamefont {Knolle},
  \citenamefont {Kovrizhin}, \citenamefont {Chalker},\ and\ \citenamefont
  {Moessner}}]{knolle15}%
  \BibitemOpen
  \bibfield  {author} {\bibinfo {author} {\bibfnamefont {J.}~\bibnamefont
  {Knolle}}, \bibinfo {author} {\bibfnamefont {D.~L.}\ \bibnamefont
  {Kovrizhin}}, \bibinfo {author} {\bibfnamefont {J.~T.}\ \bibnamefont
  {Chalker}}, \ and\ \bibinfo {author} {\bibfnamefont {R.}~\bibnamefont
  {Moessner}},\ }\href {\doibase 10.1103/PhysRevB.92.115127} {\bibfield
  {journal} {\bibinfo  {journal} {Phys. Rev. B}\ }\textbf {\bibinfo {volume}
  {92}},\ \bibinfo {pages} {115127} (\bibinfo {year} {2015})}\BibitemShut
  {NoStop}%
\bibitem [{\citenamefont {Rachel}\ \emph {et~al.}(2016)\citenamefont {Rachel},
  \citenamefont {Fritz},\ and\ \citenamefont {Vojta}}]{rachel16}%
  \BibitemOpen
  \bibfield  {author} {\bibinfo {author} {\bibfnamefont {S.}~\bibnamefont
  {Rachel}}, \bibinfo {author} {\bibfnamefont {L.}~\bibnamefont {Fritz}}, \
  and\ \bibinfo {author} {\bibfnamefont {M.}~\bibnamefont {Vojta}},\ }\href
  {\doibase 10.1103/PhysRevLett.116.167201} {\bibfield  {journal} {\bibinfo
  {journal} {Phys. Rev. Lett.}\ }\textbf {\bibinfo {volume} {116}},\ \bibinfo
  {pages} {167201} (\bibinfo {year} {2016})}\BibitemShut {NoStop}%
\bibitem [{\citenamefont {Theveniaut}\ and\ \citenamefont
  {Vojta}()}]{theveniaut17}%
  \BibitemOpen
  \bibfield  {author} {\bibinfo {author} {\bibfnamefont {H.}~\bibnamefont
  {Theveniaut}}\ and\ \bibinfo {author} {\bibfnamefont {M.}~\bibnamefont
  {Vojta}},\ }\href {https://arxiv.org/abs/1705.08913} {}\bibinfo {note}
  {ArXiv:1705.08913 (unpublished)}\BibitemShut {NoStop}%
\bibitem [{\citenamefont {Lahtinen}\ \emph {et~al.}(2012)\citenamefont
  {Lahtinen}, \citenamefont {Ludwig}, \citenamefont {Pachos},\ and\
  \citenamefont {Trebst}}]{lahtinen12}%
  \BibitemOpen
  \bibfield  {author} {\bibinfo {author} {\bibfnamefont {V.}~\bibnamefont
  {Lahtinen}}, \bibinfo {author} {\bibfnamefont {A.~W.~W.}\ \bibnamefont
  {Ludwig}}, \bibinfo {author} {\bibfnamefont {J.~K.}\ \bibnamefont {Pachos}},
  \ and\ \bibinfo {author} {\bibfnamefont {S.}~\bibnamefont {Trebst}},\ }\href
  {\doibase 10.1103/PhysRevB.86.075115} {\bibfield  {journal} {\bibinfo
  {journal} {Phys. Rev. B}\ }\textbf {\bibinfo {volume} {86}},\ \bibinfo
  {pages} {075115} (\bibinfo {year} {2012})}\BibitemShut {NoStop}%
\bibitem [{\citenamefont {Lahtinen}\ \emph {et~al.}(2014)\citenamefont
  {Lahtinen}, \citenamefont {Ludwig},\ and\ \citenamefont
  {Trebst}}]{lahtinen14}%
  \BibitemOpen
  \bibfield  {author} {\bibinfo {author} {\bibfnamefont {V.}~\bibnamefont
  {Lahtinen}}, \bibinfo {author} {\bibfnamefont {A.~W.~W.}\ \bibnamefont
  {Ludwig}}, \ and\ \bibinfo {author} {\bibfnamefont {S.}~\bibnamefont
  {Trebst}},\ }\href {\doibase 10.1103/PhysRevB.89.085121} {\bibfield
  {journal} {\bibinfo  {journal} {Phys. Rev. B}\ }\textbf {\bibinfo {volume}
  {89}},\ \bibinfo {pages} {085121} (\bibinfo {year} {2014})}\BibitemShut
  {NoStop}%
\bibitem [{\citenamefont {Wen}(2002)}]{wen02}%
  \BibitemOpen
  \bibfield  {author} {\bibinfo {author} {\bibfnamefont {X.-G.}\ \bibnamefont
  {Wen}},\ }\href {\doibase 10.1103/PhysRevB.65.165113} {\bibfield  {journal}
  {\bibinfo  {journal} {Phys. Rev. B}\ }\textbf {\bibinfo {volume} {65}},\
  \bibinfo {pages} {165113} (\bibinfo {year} {2002})}\BibitemShut {NoStop}%
\bibitem [{\citenamefont {Abrikosov}(1965)}]{abrikosov65}%
  \BibitemOpen
  \bibfield  {author} {\bibinfo {author} {\bibfnamefont {A.~A.}\ \bibnamefont
  {Abrikosov}},\ }\href@noop {} {\bibfield  {journal} {\bibinfo  {journal}
  {Physics (Long Island City, NY)}\ }\textbf {\bibinfo {volume} {2}},\ \bibinfo
  {pages} {5} (\bibinfo {year} {1965})}\BibitemShut {NoStop}%
\bibitem [{\citenamefont {Wang}\ and\ \citenamefont
  {Vishwanath}(2006)}]{wang06}%
  \BibitemOpen
  \bibfield  {author} {\bibinfo {author} {\bibfnamefont {F.}~\bibnamefont
  {Wang}}\ and\ \bibinfo {author} {\bibfnamefont {A.}~\bibnamefont
  {Vishwanath}},\ }\href {\doibase 10.1103/PhysRevB.74.174423} {\bibfield
  {journal} {\bibinfo  {journal} {Phys. Rev. B}\ }\textbf {\bibinfo {volume}
  {74}},\ \bibinfo {pages} {174423} (\bibinfo {year} {2006})}\BibitemShut
  {NoStop}%
\bibitem [{\citenamefont {Wang}(2010)}]{wang10}%
  \BibitemOpen
  \bibfield  {author} {\bibinfo {author} {\bibfnamefont {F.}~\bibnamefont
  {Wang}},\ }\href {\doibase 10.1103/PhysRevB.82.024419} {\bibfield  {journal}
  {\bibinfo  {journal} {Phys. Rev. B}\ }\textbf {\bibinfo {volume} {82}},\
  \bibinfo {pages} {024419} (\bibinfo {year} {2010})}\BibitemShut {NoStop}%
\bibitem [{\citenamefont {Messio}\ \emph {et~al.}(2013)\citenamefont {Messio},
  \citenamefont {Lhuillier},\ and\ \citenamefont {Misguich}}]{messio13}%
  \BibitemOpen
  \bibfield  {author} {\bibinfo {author} {\bibfnamefont {L.}~\bibnamefont
  {Messio}}, \bibinfo {author} {\bibfnamefont {C.}~\bibnamefont {Lhuillier}}, \
  and\ \bibinfo {author} {\bibfnamefont {G.}~\bibnamefont {Misguich}},\ }\href
  {\doibase 10.1103/PhysRevB.87.125127} {\bibfield  {journal} {\bibinfo
  {journal} {Phys. Rev. B}\ }\textbf {\bibinfo {volume} {87}},\ \bibinfo
  {pages} {125127} (\bibinfo {year} {2013})}\BibitemShut {NoStop}%
\bibitem [{\citenamefont {Chen}\ \emph {et~al.}(2012)\citenamefont {Chen},
  \citenamefont {Essin},\ and\ \citenamefont {Hermele}}]{chen12}%
  \BibitemOpen
  \bibfield  {author} {\bibinfo {author} {\bibfnamefont {G.}~\bibnamefont
  {Chen}}, \bibinfo {author} {\bibfnamefont {A.}~\bibnamefont {Essin}}, \ and\
  \bibinfo {author} {\bibfnamefont {M.}~\bibnamefont {Hermele}},\ }\href
  {\doibase 10.1103/PhysRevB.85.094418} {\bibfield  {journal} {\bibinfo
  {journal} {Phys. Rev. B}\ }\textbf {\bibinfo {volume} {85}},\ \bibinfo
  {pages} {094418} (\bibinfo {year} {2012})}\BibitemShut {NoStop}%
\bibitem [{\citenamefont {Lu}\ and\ \citenamefont {Ran}(2011)}]{lu11}%
  \BibitemOpen
  \bibfield  {author} {\bibinfo {author} {\bibfnamefont {Y.-M.}\ \bibnamefont
  {Lu}}\ and\ \bibinfo {author} {\bibfnamefont {Y.}~\bibnamefont {Ran}},\
  }\href {\doibase 10.1103/PhysRevB.84.024420} {\bibfield  {journal} {\bibinfo
  {journal} {Phys. Rev. B}\ }\textbf {\bibinfo {volume} {84}},\ \bibinfo
  {pages} {024420} (\bibinfo {year} {2011})}\BibitemShut {NoStop}%
\bibitem [{\citenamefont {Lu}\ \emph {et~al.}(2011)\citenamefont {Lu},
  \citenamefont {Ran},\ and\ \citenamefont {Lee}}]{lu11_2}%
  \BibitemOpen
  \bibfield  {author} {\bibinfo {author} {\bibfnamefont {Y.-M.}\ \bibnamefont
  {Lu}}, \bibinfo {author} {\bibfnamefont {Y.}~\bibnamefont {Ran}}, \ and\
  \bibinfo {author} {\bibfnamefont {P.~A.}\ \bibnamefont {Lee}},\ }\href
  {\doibase 10.1103/PhysRevB.83.224413} {\bibfield  {journal} {\bibinfo
  {journal} {Phys. Rev. B}\ }\textbf {\bibinfo {volume} {83}},\ \bibinfo
  {pages} {224413} (\bibinfo {year} {2011})}\BibitemShut {NoStop}%
\bibitem [{\citenamefont {Messio}\ \emph {et~al.}(2012)\citenamefont {Messio},
  \citenamefont {Bernu},\ and\ \citenamefont {Lhuillier}}]{messio12}%
  \BibitemOpen
  \bibfield  {author} {\bibinfo {author} {\bibfnamefont {L.}~\bibnamefont
  {Messio}}, \bibinfo {author} {\bibfnamefont {B.}~\bibnamefont {Bernu}}, \
  and\ \bibinfo {author} {\bibfnamefont {C.}~\bibnamefont {Lhuillier}},\ }\href
  {\doibase 10.1103/PhysRevLett.108.207204} {\bibfield  {journal} {\bibinfo
  {journal} {Phys. Rev. Lett.}\ }\textbf {\bibinfo {volume} {108}},\ \bibinfo
  {pages} {207204} (\bibinfo {year} {2012})}\BibitemShut {NoStop}%
\bibitem [{\citenamefont {Bieri}\ \emph {et~al.}(2016)\citenamefont {Bieri},
  \citenamefont {Lhuillier},\ and\ \citenamefont {Messio}}]{bieri16}%
  \BibitemOpen
  \bibfield  {author} {\bibinfo {author} {\bibfnamefont {S.}~\bibnamefont
  {Bieri}}, \bibinfo {author} {\bibfnamefont {C.}~\bibnamefont {Lhuillier}}, \
  and\ \bibinfo {author} {\bibfnamefont {L.}~\bibnamefont {Messio}},\ }\href
  {\doibase 10.1103/PhysRevB.93.094437} {\bibfield  {journal} {\bibinfo
  {journal} {Phys. Rev. B}\ }\textbf {\bibinfo {volume} {93}},\ \bibinfo
  {pages} {094437} (\bibinfo {year} {2016})}\BibitemShut {NoStop}%
\bibitem [{\citenamefont {Schaffer}\ \emph {et~al.}(2017)\citenamefont
  {Schaffer}, \citenamefont {Huh}, \citenamefont {Hwang},\ and\ \citenamefont
  {Kim}}]{schaffer17}%
  \BibitemOpen
  \bibfield  {author} {\bibinfo {author} {\bibfnamefont {R.}~\bibnamefont
  {Schaffer}}, \bibinfo {author} {\bibfnamefont {Y.}~\bibnamefont {Huh}},
  \bibinfo {author} {\bibfnamefont {K.}~\bibnamefont {Hwang}}, \ and\ \bibinfo
  {author} {\bibfnamefont {Y.~B.}\ \bibnamefont {Kim}},\ }\href {\doibase
  10.1103/PhysRevB.95.054410} {\bibfield  {journal} {\bibinfo  {journal} {Phys.
  Rev. B}\ }\textbf {\bibinfo {volume} {95}},\ \bibinfo {pages} {054410}
  (\bibinfo {year} {2017})}\BibitemShut {NoStop}%
\bibitem [{\citenamefont {Schaffer}\ \emph {et~al.}(2013)\citenamefont
  {Schaffer}, \citenamefont {Bhattacharjee},\ and\ \citenamefont
  {Kim}}]{schaffer13}%
  \BibitemOpen
  \bibfield  {author} {\bibinfo {author} {\bibfnamefont {R.}~\bibnamefont
  {Schaffer}}, \bibinfo {author} {\bibfnamefont {S.}~\bibnamefont
  {Bhattacharjee}}, \ and\ \bibinfo {author} {\bibfnamefont {Y.~B.}\
  \bibnamefont {Kim}},\ }\href {\doibase 10.1103/PhysRevB.88.174405} {\bibfield
   {journal} {\bibinfo  {journal} {Phys. Rev. B}\ }\textbf {\bibinfo {volume}
  {88}},\ \bibinfo {pages} {174405} (\bibinfo {year} {2013})}\BibitemShut
  {NoStop}%
\bibitem [{\citenamefont {Dodds}\ \emph {et~al.}(2013)\citenamefont {Dodds},
  \citenamefont {Bhattacharjee},\ and\ \citenamefont {Kim}}]{dodds13}%
  \BibitemOpen
  \bibfield  {author} {\bibinfo {author} {\bibfnamefont {T.}~\bibnamefont
  {Dodds}}, \bibinfo {author} {\bibfnamefont {S.}~\bibnamefont
  {Bhattacharjee}}, \ and\ \bibinfo {author} {\bibfnamefont {Y.~B.}\
  \bibnamefont {Kim}},\ }\href {\doibase 10.1103/PhysRevB.88.224413} {\bibfield
   {journal} {\bibinfo  {journal} {Phys. Rev. B}\ }\textbf {\bibinfo {volume}
  {88}},\ \bibinfo {pages} {224413} (\bibinfo {year} {2013})}\BibitemShut
  {NoStop}%
\bibitem [{\citenamefont {Huang}\ \emph {et~al.}(2017)\citenamefont {Huang},
  \citenamefont {Kim},\ and\ \citenamefont {Lu}}]{huang17}%
  \BibitemOpen
  \bibfield  {author} {\bibinfo {author} {\bibfnamefont {B.}~\bibnamefont
  {Huang}}, \bibinfo {author} {\bibfnamefont {Y.~B.}\ \bibnamefont {Kim}}, \
  and\ \bibinfo {author} {\bibfnamefont {Y.-M.}\ \bibnamefont {Lu}},\ }\href
  {\doibase 10.1103/PhysRevB.95.054404} {\bibfield  {journal} {\bibinfo
  {journal} {Phys. Rev. B}\ }\textbf {\bibinfo {volume} {95}},\ \bibinfo
  {pages} {054404} (\bibinfo {year} {2017})}\BibitemShut {NoStop}%
\bibitem [{\citenamefont {Reuther}\ \emph {et~al.}(2014)\citenamefont
  {Reuther}, \citenamefont {Lee},\ and\ \citenamefont {Alicea}}]{reuther14}%
  \BibitemOpen
  \bibfield  {author} {\bibinfo {author} {\bibfnamefont {J.}~\bibnamefont
  {Reuther}}, \bibinfo {author} {\bibfnamefont {S.-P.}\ \bibnamefont {Lee}}, \
  and\ \bibinfo {author} {\bibfnamefont {J.}~\bibnamefont {Alicea}},\ }\href
  {\doibase 10.1103/PhysRevB.90.174417} {\bibfield  {journal} {\bibinfo
  {journal} {Phys. Rev. B}\ }\textbf {\bibinfo {volume} {90}},\ \bibinfo
  {pages} {174417} (\bibinfo {year} {2014})}\BibitemShut {NoStop}%
\bibitem [{\citenamefont {Bernevig}\ \emph {et~al.}(2006)\citenamefont
  {Bernevig}, \citenamefont {Hughes},\ and\ \citenamefont
  {Zhang}}]{bernevig06}%
  \BibitemOpen
  \bibfield  {author} {\bibinfo {author} {\bibfnamefont {B.~A.}\ \bibnamefont
  {Bernevig}}, \bibinfo {author} {\bibfnamefont {T.~L.}\ \bibnamefont
  {Hughes}}, \ and\ \bibinfo {author} {\bibfnamefont {S.-C.}\ \bibnamefont
  {Zhang}},\ }\href {\doibase 10.1126/science.1133734} {\bibfield  {journal}
  {\bibinfo  {journal} {Science}\ }\textbf {\bibinfo {volume} {314}},\ \bibinfo
  {pages} {1757} (\bibinfo {year} {2006})}\BibitemShut {NoStop}%
\bibitem [{\citenamefont {K{\"o}nig}\ \emph {et~al.}(2008)\citenamefont
  {K{\"o}nig}, \citenamefont {Buhmann}, \citenamefont {Molenkamp},
  \citenamefont {Hughes}, \citenamefont {Liu}, \citenamefont {Qi},\ and\
  \citenamefont {Zhang}}]{konig08}%
  \BibitemOpen
  \bibfield  {author} {\bibinfo {author} {\bibfnamefont {M.}~\bibnamefont
  {K{\"o}nig}}, \bibinfo {author} {\bibfnamefont {H.}~\bibnamefont {Buhmann}},
  \bibinfo {author} {\bibfnamefont {L.~W.}\ \bibnamefont {Molenkamp}}, \bibinfo
  {author} {\bibfnamefont {T.}~\bibnamefont {Hughes}}, \bibinfo {author}
  {\bibfnamefont {C.-X.}\ \bibnamefont {Liu}}, \bibinfo {author} {\bibfnamefont
  {X.-L.}\ \bibnamefont {Qi}}, \ and\ \bibinfo {author} {\bibfnamefont {S.-C.}\
  \bibnamefont {Zhang}},\ }\href {\doibase 10.1143/JPSJ.77.031007} {\bibfield
  {journal} {\bibinfo  {journal} {J. Phys. Soc. Jpn.}\ }\textbf {\bibinfo
  {volume} {77}},\ \bibinfo {pages} {031007} (\bibinfo {year}
  {2008})}\BibitemShut {NoStop}%
\bibitem [{\citenamefont {K{\"o}nig}\ \emph {et~al.}(2007)\citenamefont
  {K{\"o}nig}, \citenamefont {Wiedmann}, \citenamefont {Br{\"u}ne},
  \citenamefont {Roth}, \citenamefont {Buhmann}, \citenamefont {Molenkamp},
  \citenamefont {Qi},\ and\ \citenamefont {Zhang}}]{konig07}%
  \BibitemOpen
  \bibfield  {author} {\bibinfo {author} {\bibfnamefont {M.}~\bibnamefont
  {K{\"o}nig}}, \bibinfo {author} {\bibfnamefont {S.}~\bibnamefont {Wiedmann}},
  \bibinfo {author} {\bibfnamefont {C.}~\bibnamefont {Br{\"u}ne}}, \bibinfo
  {author} {\bibfnamefont {A.}~\bibnamefont {Roth}}, \bibinfo {author}
  {\bibfnamefont {H.}~\bibnamefont {Buhmann}}, \bibinfo {author} {\bibfnamefont
  {L.~W.}\ \bibnamefont {Molenkamp}}, \bibinfo {author} {\bibfnamefont {X.-L.}\
  \bibnamefont {Qi}}, \ and\ \bibinfo {author} {\bibfnamefont {S.-C.}\
  \bibnamefont {Zhang}},\ }\href {\doibase 10.1126/science.1148047} {\bibfield
  {journal} {\bibinfo  {journal} {Science}\ }\textbf {\bibinfo {volume}
  {318}},\ \bibinfo {pages} {766} (\bibinfo {year} {2007})}\BibitemShut
  {NoStop}%
\bibitem [{\citenamefont {Slager}\ \emph {et~al.}(2015)\citenamefont {Slager},
  \citenamefont {Rademaker}, \citenamefont {Zaanen},\ and\ \citenamefont
  {Balents}}]{slager15}%
  \BibitemOpen
  \bibfield  {author} {\bibinfo {author} {\bibfnamefont {R.-J.}\ \bibnamefont
  {Slager}}, \bibinfo {author} {\bibfnamefont {L.}~\bibnamefont {Rademaker}},
  \bibinfo {author} {\bibfnamefont {J.}~\bibnamefont {Zaanen}}, \ and\ \bibinfo
  {author} {\bibfnamefont {L.}~\bibnamefont {Balents}},\ }\href {\doibase
  10.1103/PhysRevB.92.085126} {\bibfield  {journal} {\bibinfo  {journal} {Phys.
  Rev. B}\ }\textbf {\bibinfo {volume} {92}},\ \bibinfo {pages} {085126}
  (\bibinfo {year} {2015})}\BibitemShut {NoStop}%
\bibitem [{\citenamefont {Fu}\ and\ \citenamefont {Kane}(2007)}]{fu07}%
  \BibitemOpen
  \bibfield  {author} {\bibinfo {author} {\bibfnamefont {L.}~\bibnamefont
  {Fu}}\ and\ \bibinfo {author} {\bibfnamefont {C.~L.}\ \bibnamefont {Kane}},\
  }\href {\doibase 10.1103/PhysRevB.76.045302} {\bibfield  {journal} {\bibinfo
  {journal} {Phys. Rev. B}\ }\textbf {\bibinfo {volume} {76}},\ \bibinfo
  {pages} {045302} (\bibinfo {year} {2007})}\BibitemShut {NoStop}%
\bibitem [{\citenamefont {Fu}\ \emph {et~al.}(2007)\citenamefont {Fu},
  \citenamefont {Kane},\ and\ \citenamefont {Mele}}]{fu07_2}%
  \BibitemOpen
  \bibfield  {author} {\bibinfo {author} {\bibfnamefont {L.}~\bibnamefont
  {Fu}}, \bibinfo {author} {\bibfnamefont {C.~L.}\ \bibnamefont {Kane}}, \ and\
  \bibinfo {author} {\bibfnamefont {E.~J.}\ \bibnamefont {Mele}},\ }\href
  {\doibase 10.1103/PhysRevLett.98.106803} {\bibfield  {journal} {\bibinfo
  {journal} {Phys. Rev. Lett.}\ }\textbf {\bibinfo {volume} {98}},\ \bibinfo
  {pages} {106803} (\bibinfo {year} {2007})}\BibitemShut {NoStop}%
\bibitem [{\citenamefont {Hasan}\ and\ \citenamefont {Kane}(2010)}]{hasan10}%
  \BibitemOpen
  \bibfield  {author} {\bibinfo {author} {\bibfnamefont {M.~Z.}\ \bibnamefont
  {Hasan}}\ and\ \bibinfo {author} {\bibfnamefont {C.~L.}\ \bibnamefont
  {Kane}},\ }\href {\doibase 10.1103/RevModPhys.82.3045} {\bibfield  {journal}
  {\bibinfo  {journal} {Rev. Mod. Phys.}\ }\textbf {\bibinfo {volume} {82}},\
  \bibinfo {pages} {3045} (\bibinfo {year} {2010})}\BibitemShut {NoStop}%
\bibitem [{\citenamefont {Altland}\ and\ \citenamefont
  {Zirnbauer}(1997)}]{altland97}%
  \BibitemOpen
  \bibfield  {author} {\bibinfo {author} {\bibfnamefont {A.}~\bibnamefont
  {Altland}}\ and\ \bibinfo {author} {\bibfnamefont {M.~R.}\ \bibnamefont
  {Zirnbauer}},\ }\href {\doibase 10.1103/PhysRevB.55.1142} {\bibfield
  {journal} {\bibinfo  {journal} {Phys. Rev. B}\ }\textbf {\bibinfo {volume}
  {55}},\ \bibinfo {pages} {1142} (\bibinfo {year} {1997})}\BibitemShut
  {NoStop}%
\bibitem [{\citenamefont {Essin}\ and\ \citenamefont
  {Hermele}(2013)}]{essin13}%
  \BibitemOpen
  \bibfield  {author} {\bibinfo {author} {\bibfnamefont {A.~M.}\ \bibnamefont
  {Essin}}\ and\ \bibinfo {author} {\bibfnamefont {M.}~\bibnamefont
  {Hermele}},\ }\href {\doibase 10.1103/PhysRevB.87.104406} {\bibfield
  {journal} {\bibinfo  {journal} {Phys. Rev. B}\ }\textbf {\bibinfo {volume}
  {87}},\ \bibinfo {pages} {104406} (\bibinfo {year} {2013})}\BibitemShut
  {NoStop}%
\bibitem [{\citenamefont {Roy}(2009)}]{roy09}%
  \BibitemOpen
  \bibfield  {author} {\bibinfo {author} {\bibfnamefont {R.}~\bibnamefont
  {Roy}},\ }\href {\doibase 10.1103/PhysRevB.79.195321} {\bibfield  {journal}
  {\bibinfo  {journal} {Phys. Rev. B}\ }\textbf {\bibinfo {volume} {79}},\
  \bibinfo {pages} {195321} (\bibinfo {year} {2009})}\BibitemShut {NoStop}%
\bibitem [{\citenamefont {Qi}\ \emph {et~al.}(2009)\citenamefont {Qi},
  \citenamefont {Hughes}, \citenamefont {Raghu},\ and\ \citenamefont
  {Zhang}}]{qi09}%
  \BibitemOpen
  \bibfield  {author} {\bibinfo {author} {\bibfnamefont {X.-L.}\ \bibnamefont
  {Qi}}, \bibinfo {author} {\bibfnamefont {T.~L.}\ \bibnamefont {Hughes}},
  \bibinfo {author} {\bibfnamefont {S.}~\bibnamefont {Raghu}}, \ and\ \bibinfo
  {author} {\bibfnamefont {S.-C.}\ \bibnamefont {Zhang}},\ }\href {\doibase
  10.1103/PhysRevLett.102.187001} {\bibfield  {journal} {\bibinfo  {journal}
  {Phys. Rev. Lett.}\ }\textbf {\bibinfo {volume} {102}},\ \bibinfo {pages}
  {187001} (\bibinfo {year} {2009})}\BibitemShut {NoStop}%
\bibitem [{\citenamefont {Qi}\ \emph {et~al.}(2010)\citenamefont {Qi},
  \citenamefont {Hughes},\ and\ \citenamefont {Zhang}}]{qi10}%
  \BibitemOpen
  \bibfield  {author} {\bibinfo {author} {\bibfnamefont {X.-L.}\ \bibnamefont
  {Qi}}, \bibinfo {author} {\bibfnamefont {T.~L.}\ \bibnamefont {Hughes}}, \
  and\ \bibinfo {author} {\bibfnamefont {S.-C.}\ \bibnamefont {Zhang}},\ }\href
  {\doibase 10.1103/PhysRevB.81.134508} {\bibfield  {journal} {\bibinfo
  {journal} {Phys. Rev. B}\ }\textbf {\bibinfo {volume} {81}},\ \bibinfo
  {pages} {134508} (\bibinfo {year} {2010})}\BibitemShut {NoStop}%
\bibitem [{\citenamefont {Qi}\ and\ \citenamefont {Zhang}(2011)}]{qi11}%
  \BibitemOpen
  \bibfield  {author} {\bibinfo {author} {\bibfnamefont {X.-L.}\ \bibnamefont
  {Qi}}\ and\ \bibinfo {author} {\bibfnamefont {S.-C.}\ \bibnamefont {Zhang}},\
  }\href {\doibase 10.1103/RevModPhys.83.1057} {\bibfield  {journal} {\bibinfo
  {journal} {Rev. Mod. Phys.}\ }\textbf {\bibinfo {volume} {83}},\ \bibinfo
  {pages} {1057} (\bibinfo {year} {2011})}\BibitemShut {NoStop}%
\bibitem [{\citenamefont {Fu}(2011)}]{fu11}%
  \BibitemOpen
  \bibfield  {author} {\bibinfo {author} {\bibfnamefont {L.}~\bibnamefont
  {Fu}},\ }\href {\doibase 10.1103/PhysRevLett.106.106802} {\bibfield
  {journal} {\bibinfo  {journal} {Phys. Rev. Lett.}\ }\textbf {\bibinfo
  {volume} {106}},\ \bibinfo {pages} {106802} (\bibinfo {year}
  {2011})}\BibitemShut {NoStop}%
\bibitem [{\citenamefont {Slager}\ \emph {et~al.}(2013)\citenamefont {Slager},
  \citenamefont {Mesaros}, \citenamefont {Juricic},\ and\ \citenamefont
  {Zaanen}}]{slager13}%
  \BibitemOpen
  \bibfield  {author} {\bibinfo {author} {\bibfnamefont {R.-J.}\ \bibnamefont
  {Slager}}, \bibinfo {author} {\bibfnamefont {A.}~\bibnamefont {Mesaros}},
  \bibinfo {author} {\bibfnamefont {V.}~\bibnamefont {Juricic}}, \ and\
  \bibinfo {author} {\bibfnamefont {J.}~\bibnamefont {Zaanen}},\ }\href
  {http://www.nature.com/nphys/journal/v9/n2/abs/nphys2513.html} {\bibfield
  {journal} {\bibinfo  {journal} {Nat. Phys.}\ }\textbf {\bibinfo {volume}
  {9}},\ \bibinfo {pages} {98} (\bibinfo {year} {2013})}\BibitemShut {NoStop}%
\bibitem [{\citenamefont {Hsieh}\ \emph {et~al.}(2012)\citenamefont {Hsieh},
  \citenamefont {Lin}, \citenamefont {Liu}, \citenamefont {Duan}, \citenamefont
  {Bansil},\ and\ \citenamefont {Fu}}]{hsieh12}%
  \BibitemOpen
  \bibfield  {author} {\bibinfo {author} {\bibfnamefont {T.~H.}\ \bibnamefont
  {Hsieh}}, \bibinfo {author} {\bibfnamefont {H.}~\bibnamefont {Lin}}, \bibinfo
  {author} {\bibfnamefont {J.}~\bibnamefont {Liu}}, \bibinfo {author}
  {\bibfnamefont {W.}~\bibnamefont {Duan}}, \bibinfo {author} {\bibfnamefont
  {A.}~\bibnamefont {Bansil}}, \ and\ \bibinfo {author} {\bibfnamefont
  {L.}~\bibnamefont {Fu}},\ }\href {http://www.nature.com/articles/ncomms1969}
  {\bibfield  {journal} {\bibinfo  {journal} {Nat. Comm.}\ }\textbf {\bibinfo
  {volume} {3}},\ \bibinfo {pages} {982 EP } (\bibinfo {year}
  {2012})}\BibitemShut {NoStop}%
\bibitem [{\citenamefont {Yao}\ and\ \citenamefont {Kivelson}(2007)}]{yao07}%
  \BibitemOpen
  \bibfield  {author} {\bibinfo {author} {\bibfnamefont {H.}~\bibnamefont
  {Yao}}\ and\ \bibinfo {author} {\bibfnamefont {S.~A.}\ \bibnamefont
  {Kivelson}},\ }\href {\doibase 10.1103/PhysRevLett.99.247203} {\bibfield
  {journal} {\bibinfo  {journal} {Phys. Rev. Lett.}\ }\textbf {\bibinfo
  {volume} {99}},\ \bibinfo {pages} {247203} (\bibinfo {year}
  {2007})}\BibitemShut {NoStop}%
\bibitem [{\citenamefont {Feng}\ \emph {et~al.}(2007)\citenamefont {Feng},
  \citenamefont {Zhang},\ and\ \citenamefont {Xiang}}]{feng07}%
  \BibitemOpen
  \bibfield  {author} {\bibinfo {author} {\bibfnamefont {X.-Y.}\ \bibnamefont
  {Feng}}, \bibinfo {author} {\bibfnamefont {G.-M.}\ \bibnamefont {Zhang}}, \
  and\ \bibinfo {author} {\bibfnamefont {T.}~\bibnamefont {Xiang}},\ }\href
  {\doibase 10.1103/PhysRevLett.98.087204} {\bibfield  {journal} {\bibinfo
  {journal} {Phys. Rev. Lett.}\ }\textbf {\bibinfo {volume} {98}},\ \bibinfo
  {pages} {087204} (\bibinfo {year} {2007})}\BibitemShut {NoStop}%
\bibitem [{\citenamefont {Senthil}()}]{senthil01}%
  \BibitemOpen
  \bibfield  {author} {\bibinfo {author} {\bibfnamefont {T.}~\bibnamefont
  {Senthil}},\ }\href@noop {} {}\bibinfo {note} {ArXiv:cond-mat/0105104
  (unpublished)}\BibitemShut {NoStop}%
\bibitem [{\citenamefont {Moessner}\ \emph {et~al.}(2001)\citenamefont
  {Moessner}, \citenamefont {Sondhi},\ and\ \citenamefont
  {Fradkin}}]{moessner01}%
  \BibitemOpen
  \bibfield  {author} {\bibinfo {author} {\bibfnamefont {R.}~\bibnamefont
  {Moessner}}, \bibinfo {author} {\bibfnamefont {S.~L.}\ \bibnamefont
  {Sondhi}}, \ and\ \bibinfo {author} {\bibfnamefont {E.}~\bibnamefont
  {Fradkin}},\ }\href {\doibase 10.1103/PhysRevB.65.024504} {\bibfield
  {journal} {\bibinfo  {journal} {Phys. Rev. B}\ }\textbf {\bibinfo {volume}
  {65}},\ \bibinfo {pages} {024504} (\bibinfo {year} {2001})}\BibitemShut
  {NoStop}%
\bibitem [{\citenamefont {Misguich}\ \emph {et~al.}(2002)\citenamefont
  {Misguich}, \citenamefont {Serban},\ and\ \citenamefont
  {Pasquier}}]{misguich02}%
  \BibitemOpen
  \bibfield  {author} {\bibinfo {author} {\bibfnamefont {G.}~\bibnamefont
  {Misguich}}, \bibinfo {author} {\bibfnamefont {D.}~\bibnamefont {Serban}}, \
  and\ \bibinfo {author} {\bibfnamefont {V.}~\bibnamefont {Pasquier}},\ }\href
  {\doibase 10.1103/PhysRevLett.89.137202} {\bibfield  {journal} {\bibinfo
  {journal} {Phys. Rev. Lett.}\ }\textbf {\bibinfo {volume} {89}},\ \bibinfo
  {pages} {137202} (\bibinfo {year} {2002})}\BibitemShut {NoStop}%
\bibitem [{\citenamefont {Nikolic}\ and\ \citenamefont
  {Senthil}(2003)}]{nikolic03}%
  \BibitemOpen
  \bibfield  {author} {\bibinfo {author} {\bibfnamefont {P.}~\bibnamefont
  {Nikolic}}\ and\ \bibinfo {author} {\bibfnamefont {T.}~\bibnamefont
  {Senthil}},\ }\href {\doibase 10.1103/PhysRevB.68.214415} {\bibfield
  {journal} {\bibinfo  {journal} {Phys. Rev. B}\ }\textbf {\bibinfo {volume}
  {68}},\ \bibinfo {pages} {214415} (\bibinfo {year} {2003})}\BibitemShut
  {NoStop}%
\bibitem [{\citenamefont {Alicea}(2012)}]{alicea12}%
  \BibitemOpen
  \bibfield  {author} {\bibinfo {author} {\bibfnamefont {J.}~\bibnamefont
  {Alicea}},\ }\href {http://stacks.iop.org/0034-4885/75/i=7/a=076501}
  {\bibfield  {journal} {\bibinfo  {journal} {Reports on Progress in Physics}\
  }\textbf {\bibinfo {volume} {75}},\ \bibinfo {pages} {076501} (\bibinfo
  {year} {2012})}\BibitemShut {NoStop}%
\bibitem [{\citenamefont {Kopnin}\ and\ \citenamefont
  {Salomaa}(1991)}]{kopnin91}%
  \BibitemOpen
  \bibfield  {author} {\bibinfo {author} {\bibfnamefont {N.~B.}\ \bibnamefont
  {Kopnin}}\ and\ \bibinfo {author} {\bibfnamefont {M.~M.}\ \bibnamefont
  {Salomaa}},\ }\href {\doibase 10.1103/PhysRevB.44.9667} {\bibfield  {journal}
  {\bibinfo  {journal} {Phys. Rev. B}\ }\textbf {\bibinfo {volume} {44}},\
  \bibinfo {pages} {9667} (\bibinfo {year} {1991})}\BibitemShut {NoStop}%
\bibitem [{\citenamefont {Read}\ and\ \citenamefont {Green}(2000)}]{read00}%
  \BibitemOpen
  \bibfield  {author} {\bibinfo {author} {\bibfnamefont {N.}~\bibnamefont
  {Read}}\ and\ \bibinfo {author} {\bibfnamefont {D.}~\bibnamefont {Green}},\
  }\href {\doibase 10.1103/PhysRevB.61.10267} {\bibfield  {journal} {\bibinfo
  {journal} {Phys. Rev. B}\ }\textbf {\bibinfo {volume} {61}},\ \bibinfo
  {pages} {10267} (\bibinfo {year} {2000})}\BibitemShut {NoStop}%
\bibitem [{\citenamefont {Gurarie}\ and\ \citenamefont
  {Radzihovsky}(2007)}]{gurarie07}%
  \BibitemOpen
  \bibfield  {author} {\bibinfo {author} {\bibfnamefont {V.}~\bibnamefont
  {Gurarie}}\ and\ \bibinfo {author} {\bibfnamefont {L.}~\bibnamefont
  {Radzihovsky}},\ }\href {\doibase 10.1103/PhysRevB.75.212509} {\bibfield
  {journal} {\bibinfo  {journal} {Phys. Rev. B}\ }\textbf {\bibinfo {volume}
  {75}},\ \bibinfo {pages} {212509} (\bibinfo {year} {2007})}\BibitemShut
  {NoStop}%
\bibitem [{\citenamefont {Cheng}\ \emph {et~al.}(2009)\citenamefont {Cheng},
  \citenamefont {Lutchyn}, \citenamefont {Galitski},\ and\ \citenamefont
  {Das~Sarma}}]{cheng09}%
  \BibitemOpen
  \bibfield  {author} {\bibinfo {author} {\bibfnamefont {M.}~\bibnamefont
  {Cheng}}, \bibinfo {author} {\bibfnamefont {R.~M.}\ \bibnamefont {Lutchyn}},
  \bibinfo {author} {\bibfnamefont {V.}~\bibnamefont {Galitski}}, \ and\
  \bibinfo {author} {\bibfnamefont {S.}~\bibnamefont {Das~Sarma}},\ }\href
  {\doibase 10.1103/PhysRevLett.103.107001} {\bibfield  {journal} {\bibinfo
  {journal} {Phys. Rev. Lett.}\ }\textbf {\bibinfo {volume} {103}},\ \bibinfo
  {pages} {107001} (\bibinfo {year} {2009})}\BibitemShut {NoStop}%
\end{thebibliography}
\end{document}